# Quasi-particles models for the representations of Lie algebras and geometry of flag manifold

B. L. Feigin, A. V. Stoyanovsky

June, 1993

## 1 Introduction

Let $\mathfrak{g}$ be a finite-dimensional simply laced (we need it sometimes) semi-simple Lie algebra and $\hat{\mathfrak{g}}$ is the corresponding affine Lie algebra; $\hat{\mathfrak{g}}$ is a central extension of the current algebra $\mathfrak{g}^S$ on a circle, $\mathfrak{g}^S = \ldots + \mathfrak{g}z^{-2} + \mathfrak{g}z^{-1} + \mathfrak{g} + \mathfrak{g}z + \mathfrak{g}z^2 + \ldots$ with help of 1-dimensional space $\mathbb{C} \cdot K$, $K$ is a central element. We shall deal only with the representations with highest weight. Recall that the level of representation is the eigenvalue of $K$, we always consider the modules where $K$ acts by a scalar. It is known that $\hat{\mathfrak{g}}$ has a remarkable class of integrable representations, which can be characterized in the following way. Let $\theta$ be the root generator of $\mathfrak{g}$, corresponding to the maximal root, and put $\theta_i = \theta \cdot z^i$ and $S_i^{(k+1)} = \sum_{\alpha_1+\ldots+\alpha_{k+1}=i} \theta_{\alpha_1}\theta_{\alpha_2}\ldots\theta_{\alpha_{k+1}}$. This sum is infinite, but nevertheless $S_i^{(k+1)}$ acts in an arbitrary representation with highest weight. A representation $\pi$ of level $k$ from the category of representations with highest weight is a sum of irreducible integrable representations if and only if $k \in \mathbb{Z}, k \geq 0$ and each $S_i^{(k+1)}$, $i \in \mathbb{Z}$, is acting by zero in $\pi$. In other words, elements $S_i^{(k+1)}$ generate a two-sided ideal $T(k)$ which acts by zero on integrable representations of level k.

Let us restrict ourselves by $V_k$-the vacuum irreducible representation of level $k$. Let $v$ be the vacuum vector in $V_k$, $\hat{\mathfrak{g}}^{in} \cdot v = 0$, $\hat{\mathfrak{g}}^{in} = \mathfrak{g} + \mathfrak{g}z + \mathfrak{g}z^2 + \ldots$. Denote by $\hat{\mathfrak{g}}^-$ the algebra of creation operators $\hat{\mathfrak{g}}^- = \mathfrak{g}z^{-1} + \mathfrak{g}z^{-2} + \ldots$. Module $V_k$ is isomorphic to a quotient $U(\hat{\mathfrak{g}}^-)/I(k)$, where the submodule $I(k)$ is generated by the vector $\theta_{-1}^{k+1}v$. Algebra $U(\hat{\mathfrak{g}}^-)$ has a standard filtration such that $U(\hat{\mathfrak{g}}^-)^{ad} \cong S^*(\hat{\mathfrak{g}}^-)$. The submodule $I(k)$ defines the "space of its highest symbols" - an ideal $I(k)^{ad} \subset S^*(\hat{\mathfrak{g}}^-)$ and we get the "abealization" $V_k^{ad}$ of the representation $V_k$, $V_k^{ad} = S^*(\hat{\mathfrak{g}}^-)/I(k)^{ad}$. The dual space $(V_k^{ad})^*$ has the following nice description. Note first that the dual space $(\hat{\mathfrak{g}}^-)^*$ is naturally isomorphic to the space of 1-forms on a line $\Omega^1 \otimes \mathfrak{g}^* = F$ with values in $\mathfrak{g}^*$ (pairing is given by the residue). Coalgebra $(U(\hat{\mathfrak{g}}^-)^{ad})^* \cong S^*F$ and we realize $S^nF$ as the space of expressions $f(z_1, \ldots, z_n)dz_1 \ldots dz_n$, where $f$ is a symmetric function in $z_1, \ldots, z_n$ with values in $\mathfrak{g}^* \otimes \cdots \otimes \mathfrak{g}^*$ ($n$ times). Vector $\theta^{k+1}$ is a highest weight vector in the $\mathfrak{g}$-module $S^{k+1}\mathfrak{g}$ and let $\pi_{k+1}$ be the submodule of $S^{k+1}\mathfrak{g}$ generated by $\theta^{k+1}$, so we have an embedding $\pi_{k+1} \hookrightarrow S^{k+1}\mathfrak{g}$ and the dual map $\varphi : S^{k+1}\mathfrak{g}^* \to \pi_{k+1}^*$. Now, $(V_k^{ad})^* \subset S^*F$ consists of the elements $f(z_1 \ldots z_n)dz_1 \ldots dz_n, n \in \mathbb{Z}, n \geq 0$ which satisfy the following condition. Let $n > k+1$ and consider arbitrary point $u : (z_1, \ldots, z_n) \in \mathbb{C}^n$ such that $z_1 = z_2 = \ldots = z_{k+1}$. At this point $u$, $f(u) \in S^{k+1}(\mathfrak{g}^*) \otimes \mathfrak{g}^* \otimes \cdots \otimes \mathfrak{g}^*$ and our condition is $(\varphi \otimes 1)f(u) \in \pi_{k+1}^* \otimes \mathfrak{g}^* \otimes \cdots \otimes \mathfrak{g}^*$ is zero. The space of such functions forms an $n$-symmetric power of $F$ "with restriction to $(k+1)$-diagonal."

Let us consider more simple example of such construction. Let $T^1$ be the space of polynomial 1-forms on a line. The usual symmetric power $S^n T^1$ is realized in the space of forms $f(z_1, \ldots, z_n)dz_1 \cdots dz_n$ where $f$ is a symmetric function. The "restricted symmetric power" $S_{(l)}^n T^1$ consists of expressions $f(z_1, \ldots, z_n)dz_1 \ldots dz_n$ such that $f(z_1, \ldots, z_n)$ is zero if $z_1 = z_2 = \ldots = z_l$. It is clear that $S_{(l)}^* T^1 \subset S^* T^1$ is a commutative coalgebra.

This "restricted" symmetric co-algebra appears in the following context. Let $\mathfrak{g}$ be $sl_2$, $\{e, h, f\}$ - the standard basis in $sl_2$, $\{e\} = n$ - maximal nilpotent subalgebra in $sl_2$, $e_i = e \otimes z^i$, $\{e_i\} = \hat{n}$ – Lie algebra of currents on a circle with values in $n$, $\hat{n}_- = \{e_i\}$, $i < 0$. Define the principal subspace $W_k$ in $V_k$ as $U(\hat{n})v = U(\hat{n}_-)v$. So, $W_k \cong U(\hat{n}_-)/I$, where $I$ is an ideal. The dual co-algebra $W_k^*$ is isomorphic to $S_{(k+1)}^* T^1$. It is possible to use this result for the description of irreducible representation $V_k$ as a linear space.

Recall that in $V_k$ there is a family $v(n)$, $n \in \mathbb{Z}$ of the so-called extremal vectors. The Weyl group of $\widehat{sl}_2$ acts on $V_k$ and $\{v(n)\}$ is the orbit of vacuum vector $v = v(0)$. So, in $V_k$ we have the set of subspaces $U(\hat{n})v(m) = W_k(m)$, $W_k(m_1)$ and $W_k(m_2)$ are isomorphic and isomorphism is given by the action of some element from the Weyl group. On the other hand there is a sequence of embeddings: $\to W_k(1) \to W_k(0) \to W_k(-1) \to \ldots$ and $V_k$ is the inductive limit of this sequence. Informally, it means that it is possible to determine the "semi-infinite restricted symmetric power" of the space $\Omega^1(S^1)$ of 1-forms on the circle. The inductive limit $W_k(-\infty) \cong V_k$ is dual to the sum $\bigoplus_i S_{(k+1)}^{\frac{\infty}{2}+i}(\Omega^1(S^1)), i \in \mathbb{Z}$. In some sense those "semi-infinite restricted symmetric powers" are very close to the spaces of semi-infinite exterior forms, but their construction is more subtle and there are many open questions.

We see, that two "functional constructions" are connected with vacuum representation $V_k$. The first one represents the space dual to the adjoint space to $V_k$ as the restricted symmetric coalgebra for $\Omega^1 \otimes \mathfrak{g}^*$ - the one forms on a line with coefficients in $\mathfrak{g}^*$. The second one gives us $V_k^*$ as the semi-infinite restricted symmetric tensors of $\Omega^1(S^1) \otimes \{e^*\}$, where $e^*$ is a generator of $n^*, n \in \mathfrak{g}$. It is interesting that both construction are deduced from the structure of the annihilating ideal. The semi-infinite construction is close to the parafermionic considerations of Lepowsky, Wilson, and Primc. The character formula, which arises from this realization, coincides with the formula of Lepowsky and Primc. In §2 we investigate the semi-infinite construction for $\widehat{sl}_2$. In §4 we deal with $sl_3$. Now we explain what to do in the general case.

Fix the Cartan decomposition $\mathfrak{g} = n_1 \oplus f \oplus n_2$, and denote by $\hat{n} = \ldots + n_2 \otimes z^{-1} + n_2 + n_2 \otimes z + \ldots$, $\hat{n}_- = n_2 \otimes z^{-1} + n_2 z^{-2} + \ldots$. First we want to describe the dual space to $W_k \subset V_k$, $W_k = U(\hat{n})v$, $v$-vacuum vector in the vacuum irreducible representation of $\hat{\mathfrak{g}}$ of level $k$. To do it let us introduce the following construction. Let $A = \{A_{ij}\}$ be Cartan matrix of $\mathfrak{g}$, $\alpha_1, \ldots, \alpha_l$ – simple positive roots of $\mathfrak{g}$. We connect with this data the multigraded vector space

$$M = \oplus M_{m_1, m_2, \ldots, m_l}, m_i \in \mathbb{Z}, m_i \geq 0.$$

$$M_{m_1, m_2, \ldots, m_l} = \{f(x_1(\alpha_1), x_2(\alpha_1), \cdots, x_{m_1}(\alpha_1); x_1(\alpha_2), \ldots, x_{m_2}(\alpha_2); \cdots;$$
$$x_1(\alpha_l), \ldots, x_{m_l}(\alpha_l)) \cdot \prod_{i'<j'}(x_i(\alpha_{i'}) - x_j(\alpha_{j'}))^{-1} \prod_{i,j} dx_i(\alpha_j)\}$$

Here $f$ is a polynomial in the variables $x_i(\alpha_j)$, $f$ is symmetric with respect to each group of variables $\{x_i(\alpha_1)\}, \{x_i(\alpha_2)\}, \ldots, \{x_i(\alpha_l)\}$. Space $M$ may be considered as a component of extended symmetric power of the space $F = M_{1,0,\ldots,0} \oplus M_{0,1,0,\ldots,0} \oplus M_{0,0,\ldots,0,1}$; $M$ is extended and not "restricted" because we have the negative powers of the difference of the arguments, so $M$ is bigger than the symmetric algebra of $F$. Now let us *add* so-called "Serre relations". Introduce the subspace $\bar{M} = \oplus \bar{M}_{m_1, m_2, \ldots, m_l}$ of $M$, where $\bar{M}_{m_1, \ldots, m_l} \subset M_{m_1, m_2, \ldots, m_l}$ and elements of $\bar{M}$ are expressions in which $f$ is zero if for arbitrary $1 \leq i, j \leq l$, $i \neq j$, $x_1(\alpha_i) = x_2(\alpha_i) = \cdots = x_{-A_{ij}+1}(\alpha_i) = x_1(\alpha_j)$. We claim that $\bar{M}$ can be identified with the dual space to $U(\hat{n}_-)$. Now let us describe $W_k \cong U(\hat{n}_-)/I$. It is clear that $(W_k)^* \subset (U(\hat{n}_-))^*$, so $(W_k)^*_{m_1, m_2, \cdots, m_l} \subset \bar{M}_{m_1, m_2, \cdots, m_l}$ and element from $\bar{M}$ belongs to $(W_k)^*$ if corresponding polynomial $f$ satisfies to condition: for each $1 \leq i \leq l$ $f$ is zero if $x_1(\alpha_i) = x_2(\alpha_i) = \cdots = x_{k+1}(\alpha_i)$. Using this model we can write down the formula for the character of $W_1$. Let $L_0$ be the energy operator and consider first the case $k = 1$. Then:

$$Tr(q^{L_0})|_{W_1} = \sum_{r_1, \cdots, r_l} \frac{q^{1/2 \sum_{i,j} A_{ij} r_i r_j}}{(q)_{r_1}(q)_{r_2} \cdots (q)_{r_l}}, \quad r_i \in \mathbb{Z}, r_i \geq 0, 1 \leq i, j \leq l.$$

Here $A_{ij}$ is the Cartan matrix. For general $k$ formula has the similar form, but we have to replace $l \to l \cdot k$ and $\{A_{ij}\}$ replace to the form with the matrix $A \otimes B_k^{-1}$, where $A$ is the same Cartan matrix and $B_k$ is the symmetrization of the Cartan matrix for series

$B_k$, so $B_k = \{b_{ij}\}, 1 \leq i,j \leq k, b_{ij} = 0$ if $|i-j| > 1, b_{ij} = -1$ if $|i-j| = -1, b_{ii} = 2$ if $1 \leq i \leq k-1, b_{k,k} = 1$. Note, that formulas of such kind appeared in the articles [5],[6] where they describe the character of the space of quasi particles in the thermodynamic Bethe anzatz.

This complicated construction gives us a "functional model" for the dual space to $W_k = U(\hat{n})v \subset V_k$. Weyl group $W_{\text{aff}}$ acts on $V_k$ and let $\{v_w = w(v), w \in W_{\text{aff}}\}$ be the set of extremal vectors. Define the set of subspaces $W_k(w) = w(W_k) = U(\hat{n})v_w$. We can choose the series of elements $w_i, i \in \mathbb{Z}, i > 0$, such that $W_k(w_i) \subset W_k(w_{i+1})$ and the inductive limit of this sequence of embeddings is $V_k$. It gives some construction of $V_k$ and a character formula which we write down in §4 for $\widehat{sl}_3$.

In this paper we discuss in some detail only the $sl_2$ case. General case is much more technical and we shall write down more about it in our next paper. Note that the notion of semi-infinite restricted powers is not fully developed in this paper even in $\widehat{sl}_2$. This notion deserves a separate investigation and we plan to do that in the future.

The second topic of this article is connected with the geometry of the flag manifold for $\hat{\mathfrak{g}}$. Let $F = \hat{G}/\hat{B}$ be a flag manifold, 1 - a Shubert cell in $F$ of dimension zero and $M$ is a closure of the orbit $\hat{N} \cdot 1$, where $\hat{N} \subset \hat{G}$ is a subgroup in $\hat{G}$ which consists in the currents with values in the maximal nilpotent subgroup in $G$. Integrable irreducible representation $V_\lambda$ with highest weight $\lambda$ is realized in the space of sections of the line bundle $L_\lambda$ on $F$. The space $U(\hat{n}) \cdot v \in V_\lambda$ ($v$-vacuum vector) is dual to the space $H^0(M, L_\lambda)$. So, it is possible to use geometric arguments when we deal with $U(\hat{n})v$. In §3 in $\widehat{sl}_2$-case we use the fixed point theorem to determine the character of $H^0(M, L_\lambda)$. Manifold $M$ in the $\widehat{sl}_2$-case is non-singular and we can apply the argument similar to the case of the full flag manifold which gives us the Weyl formula for the character. So for the space $H^0(M, L_\lambda)$ we obtain two character formulas: one is a consequence of a functional realization in the space of symmetric polynomials and the second is given by the fixed point formula. If we compare two expressions, we get the Roger-Ramanudjan and Gordon identities. We also apply to $W_k$ the Demazure-like character formula and get the same result.

In §4 we discuss $\widehat{sl}_3$ case. For $\widehat{sl}_3$ the manifold $M$ is singular, so the fixed point formula becomes much more complicated. We can not write down the fixed point formula, however we conjecture a specialization of this formula $(trq^{L_0})$, which gives interesting consequences.

This paper should be viewed as an announcement of our results. We do not supply proofs of some of our statements, and in the proofs, which we give, in particular, when we work with the infinite-dimensional flag manifold we avoid certain subtle questions to emphasize ideas rather than technical points.

Let us add a few "physical words." The structure of the space $W_k$ may be described in the following way. Suppose at each point on a line we have a "particle" $\alpha_i(z)$, $\alpha_i$ is a simple root of $\mathfrak{g}$. So, the particle has coordinate $(z)$ and color $(\alpha_i), i = 1, \cdots, m$. These particles constitute an operator-algebra with the relations: $\alpha_i(z_1) \cdot \alpha_j(z_2) = (z_1 - z_2)^{c_{ij}} B_{ij}(z_1) + \cdots$, $c_{ii} = 0, c_{ij} = -1$ if $i \neq j$, $\alpha_i(z_1)^{k+1} = 0$, and "Serre relations" - the leading term in the expansion $\alpha_i(z_1)\alpha_i(z_2)\cdots\alpha_i(z_n)\alpha_j(z_{n+1}), n = -A_{ij} + 1$ is zero. Note, that this expansion starts with the term $\sqcap(z_i - z_{n+1})^{c-1}$. From this point of view it is clear that there are relations between our objects and parafermion algebras, although we do not yet fully understand them.

## 2  $\widehat{sl}_2$-case

### 2.1  Notations.

Let $e, h, f$ be the standard basic elements of the Lie algebra $sl_2$ : $e = \begin{pmatrix} 0 & 1 \\ 0 & 0 \end{pmatrix}, h = \begin{pmatrix} 1 & 0 \\ 0 & -1 \end{pmatrix}, f = \begin{pmatrix} 0 & 0 \\ 1 & 0 \end{pmatrix}$. In the Lie algebra $\widehat{sl}_2 :\cong sl_2 \otimes \mathbb{C}[t, t^{-1}] \oplus \mathbb{C}\, C$ we have the basis $f_i = f \otimes t^i, h_i = h \otimes t^i, e_i = e \otimes t^i$, and $C$, where $C$ is the central element, and the relations are:

(2.1.1)
$$\begin{aligned}
[e_i, e_j] &= [f_i, f_j] = 0 \\
[h_i, e_j] &= 2e_{i+j}; [h_i, f_j] = -2f_{i+j} \\
[e_i, f_j] &= h_{i+j} + iC\delta_{i+j,0}; [h_i, h_j] = 2iC\delta_{i+j,0}.
\end{aligned}$$

Fix the Cartan decomposition $\widehat{sl}_2 = n_+ \oplus f \oplus n_-$, where $f = \{h_0, C\}$ is Cartan subalgebra and $n_+ = \{e_i, f_i, h_i \ (i > 0)$ and $e_0\}$, $n_- = \{e_i, f_i, h_i \ (i < 0)$ and $f_0\}$ are the annihilation and creation subalgebras. The Lie algebra $\widehat{sl}_2$ has a natural grading, $\deg e_i = \deg f_i = \deg h_i = i, \deg C = 0$. We will consider only graded representations. Let $m$ be a homogeneous element of an $sl_2$-module, the number $\deg m$ is called the energy of $m$. The affine Weyl group $W_{\mathrm{aff}}$ is isomorphic to $\mathbb{Z}_2 \ltimes \mathbb{Z}$ and can be realized as a subgroup of the group of affine transformations of a line. $W_{\mathrm{aff}}$ consists of the shifts $T^n, T^n(x) = x + n, x \in \mathbb{R}, n \in \mathbb{Z}$ and reflections $S_n, n \in \mathbb{Z}, S_n(x) = n - x$. In this notations the reflection which corresponds to the root vector $f_i$ is $S_i$ and to the root vector $e_i$ is $S_{-i}$.

The weight of a vector from a graded representation of $\widehat{sl}_2$ is a triple $(m, \lambda, k)$, where $m$ is energy and $\lambda, k$ are the eigenvalues of $h_0$ and $C$. The action of the Weyl group on the weights is given by the formulas:

$$
\begin{aligned}
(2.1.2) \quad T^n(m, \lambda, k) &= (m - \lambda n - kn^2, \lambda + 2kn, k) \\
S_0(m, \lambda, k) &= (m, -\lambda, k), \\
S_n(m, \lambda, k) &= T^n S_0(m, \lambda, k) = (m + \lambda n - kn^2, -\lambda + 2kn, k).
\end{aligned}
$$

In particular the action of $T^n$ on the root vectors:

$$
(2.1.3) \qquad T^n(e_i) = e_{i-2n}, T^n(f_i) = f_{i+2n}.
$$

Recall that it is possible to introduce "a half of a sum of positive roots" $\rho = (0, 1, +2)$.

## 2.2

*Fundamental representation* of $\widehat{sl}_2$ is the irreducible representation $V$ with highest weight $\nu = (0, 0, 1)$, $V$ is a quotient of the Verma module $M_\nu$ with the vacuum vector $\bar{v}, n_+ \bar{v} = 0$, by the submodule $M_{\underline{S_0}*\nu} + M_{\underline{S_1}*\nu}$ (here the action of $w \in W_{\text{aff}}$ is defined by $\underline{w} * \nu = w(\nu + \rho) - \rho$). Singular vectors in $M_\nu$, $f_0 \bar{v}$ and $e_{-1}^2 \bar{v}$, are the highest vectors of maximal submodule in $M_\nu$, $M_{\underline{S_0}*\nu} + M_{\underline{S_1}*\nu}$. Let us denote by $v$ the image of $\bar{v}$ under the projection $M_\nu \to V$, $v$ is called the vacuum vector of $V$.

Let $\hat{n} = (e) \otimes \mathbb{C}[t, t^{-1}]$ be an abelian subalgebra of $\widehat{sl}_2$ with the basis $\{e_i\}, i \in \mathbb{Z}$. We define the *principal subspace* $W \subset V$ as $W = U(\hat{n})v$. Here $U(\hat{n})$ is an algebra of polynomials in infinite number of generators. Vacuum vector is annihilated by $e_i, i \geq 0$, so $W = \mathbb{C}[e_{-1}, e_{-2}, \ldots]v$. It means that we can identify $W$ with a quotient $\mathbb{C}[e_{-1}, e_{-2}, \ldots]/I$, where $I$ is ideal.

**Theorem 2.2.1** *Ideal $I$ is generated by the polynomials*

$$
S_k = \sum e_i e_j, i + j = k, i < 0, k < -1.
$$

Now we explain why $\{S_k\}$ belong to $I$. Recall that the Virasoro algebra is acting in the representations of $\widehat{sl}_2$ with highest weight, operator $L_i$ is defined by the formula

$$
(2.2.2) \qquad L_i = \frac{1}{2(k+2)} : \sum_{\alpha+\beta=i} (e_\alpha f_\beta + f_\alpha e_\beta + 1/2 h_\alpha h_\beta) :
$$

where $k$ is an eigenvalue of the central element.
We need now only $L_{-1}$. It is clear that

$$
(2.2.3) \qquad [L_{-1}, e_i] = ie_{i-1}, [L_{-1}, f_i] = if_{i-1}, [L_{-1}, h_i] = ih_{i-1}
$$

Vacuum vector in the fundamental representation is annihilated by $L_{-1}, L_{-1}v = 0$. We know, that in $V$, $e_{-1}^2 v = 0$, so $(L_{-1})^p e_{-1}^2(v) = 0$, but $(L_{-1})^p (e_{-1}^2)(\bar{v}) = aS_{-2-p}(\bar{v})$, where $a \neq 0$ is some constant.

*Remark* 2.2.4 Let us write down the infinite sums $\bar{S}_m = \sum_{i+j=m} e_i e_j, m \in \mathbb{Z}, i, j \in \mathbb{Z}$. It is easy to see that $\bar{S}_m$ are well-defined operator in arbitrary representation from the category of representations with highest weight. The well-known result from the conformal field theory is that $\bar{S}_m$ acts by zero on an arbitrary integrable representation of level 1. In terms of generating function we can rewrite it as $e(z)^2 = 0$, where $e(z) = \sum_{p \in \mathbb{Z}} e_p z^p$.

## 2.3

The space $W$ is a sum of its weight subspaces : $W = \bigoplus_{n,\lambda} W_{(n,\lambda,1)}$. Our aim is to determine the character of $W$, $ch(W) \stackrel{\text{def}}{=} \Sigma q^i z^j \dim W_{(-i,2j,1)}$. First we introduce some convenient description of the dual space $W^*$.

Let us identify $\hat{n} = \{e_i\} = (e) \otimes \mathbb{C}[t, t^{-1}]$ with $\mathbb{C}[t, t^{-1}]$ -the space of algebraic functions on $\mathbb{C}^*$. Decompose $\hat{n} = \hat{n}_- \oplus \hat{n}_+, \hat{n}_- = \{e_i\}, i < 0, \hat{n}_+ = \{e_i\}, i \geq 0$, so $\hat{n}_+$ consists of functions which are regular at zero, and $\hat{n}_-$ is isomorphic to $\mathbb{C}[t, t^{-1}]/\mathbb{C}[t]$. The dual space $\hat{n}_-^* \cong (\mathbb{C}[t, t^{-1}]/\mathbb{C}[t])^*$ is naturally isomorphic to the space of polynomial 1-forms on $\mathbb{C}, \hat{n}^* \cong \Omega^1(\mathbb{C}) = \{f(x)dx, f(x) \in \mathbb{C}[x]\}, \deg x^n dx = n+1$. Therefore $(U(\hat{n}_-))^* \cong \bigoplus_{p \geq 0} S^p(\hat{n}_-)^* \cong \bigoplus_{p \geq 0} S^p(\hat{n}_-^*) \cong \bigoplus_{p \geq 0} S^p(\Omega^1(\mathbb{C}))$, where $S^p(\Omega^1(\mathbb{C}))$ is the space of expressions $f(x_1, \ldots, x_p)dx_1 dx_2 \ldots dx_p = \omega$. Note, that $\omega$ is not a volume form, here $dx_i$ and $dx_j$ commute with each other, and $f$ is a symmetric polynomial. We will call $S^p(\Omega^1(\mathbb{C}))$ the *p*-particles space. Pairing of $f dx_1 dx_i \ldots dx_p$ and the product of currents $(\varphi_1 \otimes e) \cdot (\varphi_2 \otimes e) \ldots (\varphi_p \otimes e)$ is given by the formula:

$$(2.3.1) \qquad \langle f dx_1 dx_2 \ldots dx_p, (\varphi_1 \otimes e) \ldots (\varphi_p \otimes e) \rangle$$
$$= \operatorname{Res}_{x_1 = \ldots = x_p = 0} f(x_1 \ldots x_p) \varphi_1(x_1) \ldots \varphi_p(x_p) dx_1 \ldots dx_p$$

(Res is the coefficient before the term $x_1^{-1} \ldots x_p^{-1} dx_1 \ldots dx_p$ in Loran expansion).

$W$ is a quotient $U(\hat{n}_-)/I$, so $W^*$ is a subspace in the coalgebra $U(\hat{n}_-)^*$; $W^* = \bigoplus_{p \geq 0} W_p^*, W_p^* \subset S^p(\Omega^1(\mathbb{C}))$. Using the description of generators of $I$ we obtain that $W_p^* = \{f(x_1, \ldots, x_p)dx_1 \ldots dx_p\}$ such that $f(x_1, \ldots, x_p) = 0$ when $x_1 = x_2$, in other words $f = g \cdot \prod_{i<j}(x_i - x_j)^2$. Thus

$$(2.3.2) \qquad W^* = \bigoplus_{p=0}^{\infty} W_p^*, \quad \text{where} \quad W_p^* = \{g(x_1 \ldots x_p) \prod_{i<j}(x_i - x_j)^2 \prod_{i=j}^{p} dx_i\},$$

*g*-symmetric polynomial.

Informally this picture describes the situation of nondistinguished particles, which can move on a line and two of them can not be simultaneously in one point.

Now it is easy to write down a formula for the character of $W$:

$$\text{(2.3.3)} \quad chW = chW^* = \sum_{p=0}^{\infty} chW_p^*$$

$$= \sum_{p=0}^{\infty} \frac{q^{p^2} z^p}{(1-q)(1-q^2)\ldots(1-q^p)}.$$

If $z = 1$ or $z = q$ we get the left hand sides of the famous Roger-Ramanujan identities.

## 2.4

Theorem 2.2.1 can be used for describing the whole space of the fundamental representation $V$. Let $n \in \mathbb{Z}$, $\{v_n = T^n v\}$ be the set of all extremal vectors in $V$. Due to (2.1.3), the weight of $v_n$ is equal to $(-n^2, 2n, 1)$ Consider the spaces $W_p = T^p W = U(\hat{n})v_p$. It is evident that $\ldots W_2 \subset W_1 \subset W_0 \subset W_{-1} \subset W_{-2} \subset \ldots$ (pic. 1)

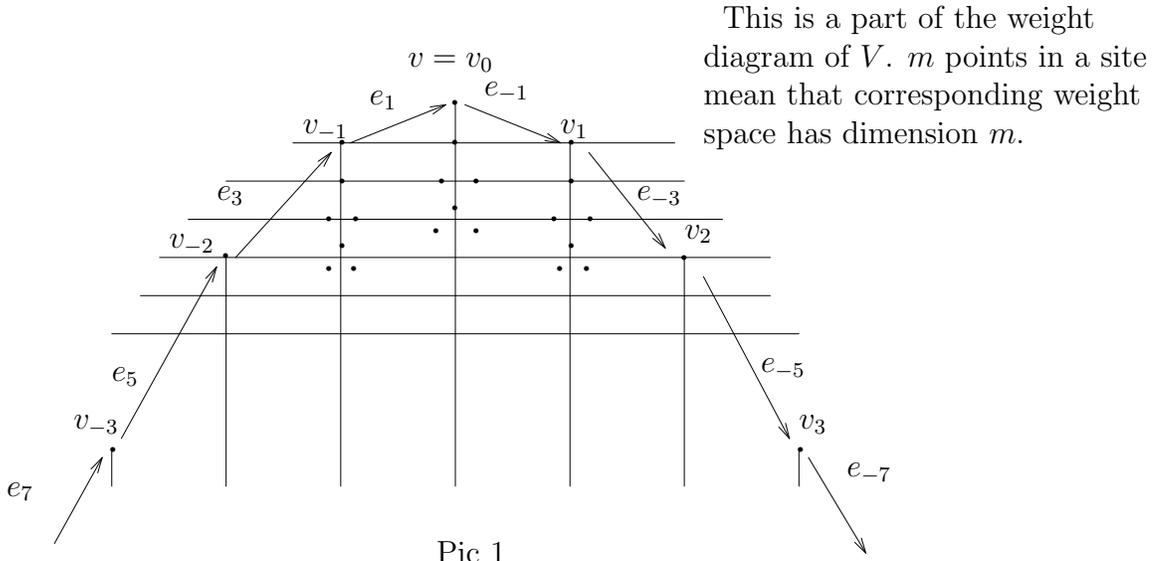

This is a part of the weight diagram of $V$. $m$ points in a site mean that corresponding weight space has dimension $m$.

Pic 1

Using the theorem 2.2.1 and formula (2.1.3), we obtain that
$W_n = (\mathbb{C}[e_{-2n-1}, e_{-2n-2}, \ldots]/I_n)v_n$, where $I_n$ is ideal, generated by the polynomial $S_m = \sum_{\alpha+\beta=m} e_\alpha e_\beta, m \geq -4n-2, \alpha, \beta \geq -2n-1$. So, we have a sequence of $\mathbb{C}[e_i], i \in \mathbb{Z}$-modules
$\ldots \to W_2 \xrightarrow{\theta_2} W_1 \xrightarrow{\theta_1} W_0 \xrightarrow{\theta_0} W_{-1} \to \ldots$, where each arrow is a $\mathbb{C}[e_i]$-homomorphism. (Note, that in $W_n, e_i, i < -2n-1$, act by zero). The formula for the embedding can be deduced from picture 1. For example: $v_0$ is the generator of $W_0$ and $\theta_0(v_0) = e_1 v_{-1}$, where $v_{-1}$ is the generator of $W_{-1}$. Our fundamental representation is an inductive limit of this sequence. This fact can be formulated in the following way. Any vector of the space $V$ is represented as a finite linear combination of vectors $e_{i_1} e_{i_2} \ldots e_{i_n} v_{-N} = e_{i_1} e_{i_2} \ldots e_{i_n} e_{2N+1} v_{-N-1} = \ldots$ if $N$ is large enough. Now let $N$ go to infinity, it means that we add (formally) "extremal" vector $v_{-\infty}$ and write down $v_{-N} = (e_{2N+1}e_{2N+3}e_{2N+5}\ldots)v_{-\infty}$.

**Theorem 2.4.1** *Let $\tilde{V}$ be the vector space with the basis of infinite monomials $M = (e_{i_1}e_{i_2}e_{i_3}\ldots)v_{-\infty}$, where the infinite sequence of indices $\{i_1, i_2, i_3, \ldots\}$ stabilizes. It means that for some $n$, $i_n$ is an odd number and $i_{p+1} = i_p + 2$, if $p \geq n$. We also suppose that (1) different $e_{i_\alpha}$ commute : $e_{i_1}e_{i_2}\ldots e_{i_k}\ldots e_{i_e}\ldots = e_{i_1}e_{i_2}\ldots e_{i_e}\ldots e_{i_k}\ldots$ and (2) if $m$ contains a fragment $e_u e_{2N+1} e_{2N+3} \ldots$ such that $u \geq 2N$ then $m = 0$. The space $\tilde{V}$ is a $\mathbb{C}[e_i] (i \in \mathbb{Z})$ module and a natural completion of algebra $\mathbb{C}[e_i]$ acts on $\tilde{V}$. In particular, the expressions $S_p = \Sigma e_\alpha e_\beta$, $p \in \mathbb{Z}$, $\alpha + \beta = p$, $\alpha, \beta \in \mathbb{Z}$ act on $\tilde{V}$. Let $V$ be the quotient $\tilde{V}/(S_p)\tilde{V}$. Then in $V$ it is possible to define an action of $\widehat{sl}_2$ such that we get the fundamental representation, where $\{e_i\}$ act just by multiplication.*

*Remark* 2.4.2 Extremal vector $v_N$ is annihilated by subalgebra $T^N n_+ T^{-N} = n_+(N) \subset \widehat{sl}_2$. If $N$ goes to minus infinity, then subalgebra $n_+(N)$ tends to $n_+(-\infty) = \{f_i, i \in \mathbb{Z}, h_i, i > 0\}$. So, it is natural to think that vector $v_{-\infty}$ is killed by subalgebra $n_+(-\infty)$. But actually vector $v_{-\infty}$ is annihilated by subalgebra $\hat{b} = \{f_i, h_i, i \in \mathbb{Z}\}$. We try to explain this point of view in the next remark.

*Remark* 2.4.3 Let us reformulate the statement of the theorem 2.4.1. in a more geometrical way. Let $\mathbb{C}P^1$ be a projective line and $\widehat{\mathbb{C}P^1}$-set of current $S^1 \to \mathbb{C}P^1$. This $\widehat{\mathbb{C}P^1}$ can be considered as infinite dimensional complex manifold. Group $SL_2^S$ acts on $\widehat{\mathbb{C}P^1}$. Fix a Shubert decomposition of $\mathbb{C}P^1 = \{p \cap \mathbb{C}\}$ and let $\hat{\mathbb{C}}$ be the space of maps $S^1 \to \mathbb{C}$, $\hat{\mathbb{C}}$ is a dense set in $\widehat{\mathbb{C}P^1}$. Denote by $\hat{\mathbb{C}}(n)$ a set of maps $S^1 \to \mathbb{C}$ which are admit analytic continuation inside the disk $D = \{z, |z| \leq 1\}$ (we identify $S^1$ with the boundary of $D$) with the possible singularity only at zero, and such that the corresponding function $f : D \to \mathbb{C}$ has the form $g(z) \cdot z^n$, where $g$ is regular at zero. The Lie algebra $\{e_i\}$ acts on $\hat{\mathbb{C}}$ in a simple way, a vector field corresponding to $e_i$ is a shift, if $f \in \hat{\mathbb{C}}, \delta f = z^i$. Now we will construct some subspace in the space of distributions on $\hat{\mathbb{C}}$. Let $\delta(n)$ be the $\delta$-function with support on $\hat{\mathbb{C}}(n)$. Informally, $\delta(n)$ is zero outside $\hat{\mathbb{C}}(n)$ and infinite on $\hat{\mathbb{C}}(n)$. Up to a constant $\delta(n)$ is characterized by two condition (1) support $\delta(n) = \hat{\mathbb{C}}(n)$ and (2) $X\delta(n) = 0$ if $X$ is vector field on $\hat{\mathbb{C}}$ tangent to $\hat{\mathbb{C}}(n)$. Introduce the space $U_2 = \bigoplus_p \mathbb{C}[e_i]\delta(p), p \in \mathbb{Z}$, $U_2$ consists of all derivations of distributions $\delta(p)$. Then let us factorize $U_2$ by the $\mathbb{C}[e_i]$ submodule which is generated by elements $\delta(p) - e_{p+1}\delta(p+2)$. We get the space $U_1$ and then factorize it by the action of elements $\Sigma e_\alpha e_\beta$. The result is our space $V$. So, this construction is very similar to the construction of the induced representation of algebra $sl_2^S$ from trivial representation of $\hat{b}$. The difference is that we work with the space of polynomials of infinite degree on the homogeneous space $\widehat{\mathbb{C}P^1}$. This infinite effects create a non-trivial central charge.

*Remark* 2.4.4 Here we try to explain the problems which arise when we try to define an action of the operators $h_i$ and $f_i$ in $V$. We know that $h_i(v_p) = 0$ if $i > 0$. It means that we can write down:

$$h_\alpha(e_{i_1} e_{i_2} \ldots v_{-\infty}) = [h_\alpha e_{i_1}]e_{i_2}\ldots v_{-\infty} + e_{i_1}[h_\alpha e_{i_2}]\ldots v_{-\infty} + \ldots .$$

But in this infinite sum only finite number of non-zero terms. The action of $h_0$, and the action of the energy operator can be found in the following way.

$$h_0(e_{2N+1}e_{2N+3}\ldots v_{-\infty}) = -2Ne_{2N+1}e_{2N+3}\ldots v_{-\infty}; h_0(e_i w) = e_i h_0(w) + 2e_i w$$
$$\deg(e_{2N+1}e_{2N+3}\ldots v_{-\infty}) = -N^2, \deg(e_i w) = \deg w + i.$$

Now let us try to find $h_{-1}(e_1 e_3 e_5 \ldots v_\infty)$.

$$\begin{aligned} h_{-1}(e_1 e_3 e_5 \ldots v_\infty) &= [h_{-1}e_1]e_3 e_5 \ldots v_{-\infty} + e_1 h_{-1} e_3 e_5 \ldots v_{-\infty} \\ &= 2e_0 e_3 e_5 \ldots v_{-\infty} + 2e_1 e_2 e_5 e_7 \ldots v_{-\infty} \\ &\quad + 2e_1 e_3 e_4 e_7 e_9 \ldots v_{-\infty} + \ldots + e_1 e_3 e_5 \ldots h_{-1} v_{-\infty}. \end{aligned}$$

Due to the previous remark, we put $h_{-1}v_{-\infty} = 0$. Using the quadratic relations, we obtain:

$$e_0 e_3 e_5 \ldots v_{-\infty} = -e_1 e_2 e_5 e_7 \ldots v_{-\infty} = e_1 e_3 e_4 e_7 e_9 \ldots v_{-\infty} = \ldots$$

So if we put $2e_0 e_3 e_5 \ldots v_{-\infty} = a$, then:

$$h_{-1}(e_1 e_3 e_5 \ldots v_{-\infty}) = (a - a + a - a + a - \ldots)e_0 e_3 e_5 \ldots v_\infty$$

The partial sums of this row are $a, 0, a, 0, \ldots$. Therefore the sum of our sequence is $\frac{a}{2}$ (as in the textbook by Hardy). We get then:

$h_{-1}(e_1 e_3 e_5 \ldots v_{-\infty}) = e_0 e_3 e_5 \ldots v_{-\infty}$. It is possible to see that $h_{-1}$ actually acts on the fundamental representation by this way. In principle it takes some work with infinites to define an action of all $h_i$, but we failed to invent the general procedure. Note, that the action of $f_i$ can be found if we know the action of $h_i$. It is clear that, $f_i(v_N) = 0$ if $N$ is small enough. Then,

$$f_\alpha(e_{i_1} e_{i_2} \ldots v_{-\infty}) = [f_\alpha e_{i_1}]e_{i_2} \ldots v_{-\infty} + e_{i_1}[f_\alpha e_{i_2}] \ldots + \ldots + e_{i_1} e_{i_2} \ldots f_\alpha v_{-\infty}.$$

The last term is zero as $f_\alpha v_{-\infty} = 0$ and only finite number of terms in the sum is non-zero.

## 2.5

At this point we present the way of constucting the operators $h_i, f_i$

**Lemma 2.5.1** *Denote be $\hat{n}(r)$ the subalgebra in $\hat{n}$ with the basis $\{e_i, i \geq r\}$. Then $U(\hat{n}(r))v_{-\infty} = V$.*

We have to explain the meaning of the symbol $U(\hat{n}(r))v_{-\infty}$. This space consists of expressions $e_{i_1}e_{i_2}e_{i_3}\ldots v_{-\infty}$, where all $i_s \geq r$. Now suppose we have an arbitrary word $e_{j_1}e_{j_2}\ldots e_{2N+1}e_{2N+3}\ldots v_{-\infty}$, where $j_1 \leq j_2 \leq \cdots$.

Using the quadratic relations, let us write down:

$$(e_{j_1}e_{2N+1} + e_{j_1+1}e_{2N} + e_{j_1+2}e_{2N-1} + \cdots + e_{2N+1}e_{j_1}) \times$$
$$\times e_{j_2}e_{j_3}\cdots e_{2N+3}e_{2N+5}\cdots v_{-\infty} = 0.$$

Therefore it is possible to express $e_{j_1}e_{j_2}\cdots e_{2N+1}e_{2N+3}\cdots v_{\infty}$ as a linear combinations of expressions "without $j_1$". We can repeat this procedure several times and replace our element by the sum of words, where $j_1, j_2, \ldots, j_M$ does not appear. It is evident, that if $M$ is big enough we got an element from $U(\hat{n}(r))v_{-\infty}$.

Now define an action of $f_\alpha$:

$$f_\alpha(e_{j_1}e_{j_2}\cdots v_{-\infty}) = -h_{\alpha+j_1}e_{j_2}e_{j_3}\cdots v_{-\infty} - e_{j_1}h_{\alpha+j_2}e_{j_3}e_{j_4}\cdots v_{-\infty} + \cdots.$$

In general case this sum is infinite, but suppose, that for all $s, \alpha + j_s > 0$. In this case only finite number of terms are non-zero and in the remark 2.4.4 we defined an action of $h_\beta$, $\beta > 0$. So we get an action of $f_\alpha$ in $V$. It is possible to verify that our definition is correct: namely if we represent the vector $w \in V$ as an element of $U(\hat{n}(r))v_{-\infty}$ where $r + \alpha > 0$ by two different way and calculate an action of $f_\alpha$ the results will be the same. Operators $h_\alpha$ are the brackets of $e_u$ and $f_v$ $u + v = \alpha$, so we can define the action of them.

## 2.6

Here we construct in the principal space $W$ and in the space of fundamental representation the monomial bases. We shall say that monomial $e_{i_1}e_{i_2}\cdots e_{i_s}v \in W$ is *basic* if $i_p + 1 < i_{p+1}$. Similarly the infinite monomial $e_{i_1}e_{i_2}\cdots v_{-\infty}$ basic if the same condition $i_p + 1 < i_{p+1}$ is hold.

**Proposition 2.6.1** *Basic monomials constitute bases in $W$ and $V$.*

This statement about $W$ is a direct consequence of Theorem 2.2.1 and $V$ is a sum of spaces $T^n W$, $n \in \mathbb{Z}$. It gives the statement about basis in $V$.

From 2.6.1 it is possible to obtain the character formula (left hand side of the Roger-Ramanujan identity) for $W$. Now we use 2.6.1 to write down a character formula for the fundamental representation. As $V = \bigcup T^n W$, $n \in \mathbb{Z}$, $ch\, V = \lim_{N \to \infty} ch\, W_{-N}$. We shall use standard notations:

$$(q)_k = (1-q)(1-q^2)\cdots(1-q^k), (q)_\infty = \prod_{i=1}^{\infty}(1-q^i),$$

$$ch(T^{-N}W) = T^{-N} ch\, W$$

$$= \sum_{k=0}^{\infty} \frac{q^{k^2}}{(q)_k}(q^{-2Nk}z^k)q^{N^2}z^{-N} \quad ((2.3.3), (2.1.2))$$

$$= \sum_{n=-N}^{\infty} q^{n^2}z^n (q)_{n+N}^{-1} \quad \text{(here we substitute } k - N = n\text{)}.$$

Then $N$ goes to infinity:

(2.6.2) $$ch\, V = \lim_{N \to \infty} ch\, W_{-N} = \frac{\sum_{n=-\infty}^{\infty} q^{n^2} z^n}{(q)_\infty}$$

This formula coincides with the well known expression for the character of fundamental representation in bosonic realization.

*Remark 2.6.3 Proposition 2.6.1 gives an expressions for the characters of $W$ and $V$ as a statistical sum of configurations of points on a 1-dimensional lattice. For $W$ it is known, and we shall formulate this fact for $V$. A configuration is an infinite set of points on a 1-dimensional lattice $i_1 < i_2 < i_3 < \cdots$ which satisfies two conditions: (1) $i_\alpha + 1 < i_{\alpha+1}$ and (2) for some $N$ $i_N$ is odd number and $i_\alpha + 2 = i_{\alpha+1}$ if $\alpha \geq N$. The statistical weight of configuration is $q^{S(A)}$, $q \in \mathbb{C}^*$, $S(A) = i_1 + i_2 + \cdots$. This definition does not make sense of because $S = \infty$. But we let us change the definition of $S(A)$. The configuration $U = \{1, 3, 5\ldots\}$ will be called vacuum. Then $S(A)$ and $S(U)$ are infinite, but the difference $S(A) - S(U)$ is of finite value. So, we define the statistical sum $\sum(q)$ as $\sum q^{S(A)-S(U)}$, where the sum is taken over all configuration. According to Proposition 2.6.1, $\sum(q)$ is just the character of the fundamental representation.*

## 2.7

The results of §§2–6 can be generalized to the case of an arbitrary integrable representation of $\widehat{sl}_2$. First we will explain what to do with the representation with highest weight $(0,1,1)$ (this representation has level 1 and is also called fundamental). In this case we also have an extremal vector "at infinity" $v_{-\infty}$, and the monomials $e_{i_1} e_{i_2} \cdots v_{-\infty}$ constitute the basis if $i_\alpha + 1 < i_{\alpha+1}$ and for some $N$ $i_N$ is an *even* number and $i_{\alpha+1} = i_\alpha + 2$ if $\alpha \geq N$. From the point of view of configurations on 1-dimensional lattice is means that configuration stabilize on "another" vacuum, which consists of the even numbers. Now let us consider the case of an arbitrary integrable representation.

Let $V$ be the irreducible representation of $\widehat{sl}_2$ with highest weight $\lambda = (0, l, k)$, where $k, l \in \mathbb{Z}$, $0 \leq l \leq k$, $v$-vacuum vector and

$$W = U(\hat{n})v = (\mathbb{C}[e_{-1}, e_{-2}, \ldots]/I_{l,k})v.$$

**Theorem 2.2.1'** *The ideal $I_{l,k}$ is generated by the polynomials $e_{-1}^{k+1-l}$ and $S_i^{(k+1)} = \sum e_{\alpha_1} e_{\alpha_2} \ldots e_{\alpha_{k+1}}$, $\alpha_j \leq -1$, $i \leq -(k+1)$.*

*Remark* Actually the following fact is true. Let $M$ be an arbitrary integrable representation of level $k$. Then the elements $S_m = \sum e_{\alpha_1} e_{\alpha_2} \ldots e_{\alpha_{k+1}}$, $\alpha_j \in \mathbb{Z}$, $\alpha_1 + \alpha_2 + \cdots + \alpha_{k+1} = m$ acts by zero on $M$. The converse is also true: if $N$ is representation of level $k$ from category of representations with highest weight and elements $S_m$ from the universal enveloping algebra act by zero on $N$, then $N$ is a direct sum of integrable representations. The annihilating ideal of all integrable modules of level $k$ is generated by $S_m$. In terms of generating functions the conditions $S_m = 0$ ($m \in \mathbb{Z}$) can be written as $e(z)^{k+1} = 0$, $e(z) = \sum e_i z^i$.

As in the case $k = 1$ the dual space $W^*$ is described in terms of the space of symmetric functions. The following generalization of (2.3.2) is true.

(2.3.2')
$$\begin{aligned} W^* &= \bigoplus_{m=0}^{\infty} W_m^*, \quad \text{where} \\ W_m^* &\cong \{f(x_1, \ldots, x_m) dx_1 dx_2 \ldots dx_m\}, \end{aligned}$$

function $f$ is symmetrical polynomial in $m$ variables $f = 0$ if $x_1 = x_2 = \cdots = x_{k+1}$ and $x_1 = x_2 = \cdots = x_{k-l+1} = 0$ (the first makes sense, if $k + 1 \leq m$, and the second, if $k - l + 1 \leq m$). In other words, $f$ is zero if $k + 1$ points coincide or $k - l + 1$ points are equal to zero. The character of $W$ is given by the left hand side of Gordon identity.

(2.3.3')
$$\begin{aligned} ch\, W &= \sum_{m=0}^{\infty} ch\, W_m^* \\ &= \sum_{m=0}^{\infty} \sum_{\substack{N_1 \geq N_2 \geq \cdots \geq N_k \geq 0 \\ N_1 + N_2 + \cdots + N_k = m}} \cdot \frac{z^{m+l/2} q^{N_1^2 + N_2^2 + \cdots + N_k^2 + N_{k-l+1} + N_{k-l+2} + \cdots + N_k}}{(q)_{N_1 - N_2}(q)_{N_2 - N_3} \cdots (q)_{N_{k-1} - N_k}(q)_{N_k}}. \end{aligned}$$

We shall give the sketch of a proof of this fact. For simplicity we restrict ourselves by the case $l = 0$.

**Theorem 2.7.1** *The character of the space* $W_m^* = \{f(x_1, \ldots, x_m)dx_1 \ldots dx_m\}$, *where f-symmetric polynomial, $f = 0$ if $x_1 = x_2 = \cdots = x_{k+1}$ is equal*

$$ch\, W_m^* = \sum_{\substack{N_1 \geq N_2 \geq \cdots \geq N_k \geq 0 \\ \sum N_i = m}} \frac{q^{N_1^2 + N_2^2 + \cdots + N_k^2}}{(q)_{N_1-N_2}(q)_{N_2-N_3}\cdots(q)_{N_{k-1}-N_k}(q)_{N_k}}.$$

Let $p = (p_1, \ldots, p_s)$ be a decomposition $m = p_1 + p_2 + \cdots + p_s$, $p_1 \geq p_2 \geq \cdots \geq p_s > 0$, $p_i \in \mathbb{Z}$. We associate to $(p)$ a subspace $U_p \subset W_m^*$ in the following way:

$$f(x_1 \ldots x_m)dx_1 \ldots dx_m \in U_p, \quad \text{if} \quad f = 0 \quad \text{when} \quad x_1 = x_2 = \cdots = x_{p_1},$$

$$x_{p_1+1} = \cdots = x_{p_1+p_2}, x_{p_1+p_2+1} = \cdots = x_{p_1+p_2+p_3}, \cdots,$$

$$x_{p_1+\cdots+p_{s-1}+1} = \cdots = x_m.$$

Define the space $\Gamma_p$ as $\cap_{p' \geq p} U_{p'}$, where $p' = (p'_1, p'_2, \ldots, p'_s)$ is greater $(p' \geq p)$ then $p$, if $p'_1 > p_1$ or $p'_1 = p_1$ and $p'_2 > p_2$, or $p'_1 = p_1$, $p'_2 = p_2$, $p'_3 > p_3, \ldots$ (this is lexicographic ordering). It is clear that $W_m^* = U_{p(k)} = \Gamma_{p(k)}$, where $p(k) = (k+1, 1, 1, \ldots, 1)$. Thus we have a filtration in $W_m^*$ which are labeled by the elements of the ordering set and let us form the adjoint graded space $Gr\,\Gamma = \oplus_p (Gr\,\Gamma)_p$. Each component $(Gr\,\Gamma)_p$ of this space can be identified with the space of expressions

$$\varphi(z_1, \ldots, z_s)(dz_1)^{p_1}(dz_2)^{p_2} \ldots (dz_s)^{p_s} \quad \text{where}$$

$$z_1 = x_1 = x_2 = \cdots = x_{p_1}, z_2 = x_{p_1+1} = \cdots = x_{p_1+p_2}, \ldots, z_s =$$

$$x_{p_1+\cdots+p_{s-1}+1} = \cdots = x_m.$$

(We have $s$ groups of clusters, each cluster is a "composite particle".) The function $\varphi$ satisfies the conditions (1) if $p_i = p_j$ then $\varphi$ is symmetric with respect to the transposition of the coordinates $z_i$ and $z_j$, (2) $\varphi$ has a zero on the diagonal $z_i = z_j$ of degree $æ_{ij}$. It is easy to see that $æ_{ij} = 2p_j$ if $p_i \geq p_j$. Namely, suppose that $p = (p_1, p_2)$. So we have variables $(x_1 \cdots x_{p_1}, x_{p_1+1}, \ldots, x_{p_1+p_2})$. The function of lowest degree from $\Gamma_p$ has a form $\text{Symm} \prod_{t=1}^{p_2}(x_t - x_{p_2+t})^2$ and the degree of if is equal $2p_2$. Our statements deduced from this. Finally:

$$(Gr\,\Gamma)_p = \{\varphi(z_1 \ldots z_s)(dz_1)^{p_1} \cdots (dz_s)^{p_s} \prod_{i<j}(z_i - z_j)^{2p_j}\},$$

where $\varphi$ satisfies the symmetry condition (1).

Let $n_r$ be a number of $p_i$ in $p = (p_1, \ldots, p_s)$ which are equal to $r$. Write down the character of $(Gr\,\Gamma)_p$:

$$ch(Gr\,\Gamma)_p = \frac{q^{\sum_{i<j} 2p_j + \sum_i p_i}}{\prod_r (q)_{n_r}} = \frac{q^{\sum_r r(n_r^2 - n_r) + \sum_{r<t} 2rn_r n_t + \sum_r rn_r}}{\prod_r (q)_{n_r}}$$

$$= \frac{q^{\sum_r (n_r + n_{r+1} + n_{r+2} + \cdots)^2}}{\prod_r (q)_{n_r}} = \frac{q^{\sum_r N_r^2}}{(q)_{N_1 - N_2}(q)_{N_2 - N_3} \cdots (q)_{N_l}}.$$

It is clear that Young diagram which corresponds to $(N_1, N_2, \ldots)$ is dual to the Young diagram which corresponds to $p$. Collect together all $ch(Gr\,\Gamma_p)$. We get:

$$ch\,S^m \Omega^1(\mathbb{C}) = \frac{q^m}{(1-q)\cdots(1-q^m)}$$

$$= \sum_{\substack{(N_1, N_2, \ldots) \\ N_1 \geq N_2 \geq \cdots \geq 0, \sum N_i = m}} \frac{q^{\sum N_r^2}}{(q)_{N_1 - N_2}(q)_{N_2 - N_3} \cdots}.$$

The character of $W_m^*$ is a sum of $ch(Gr\,\Gamma)_p$ such that $p < p(k)$. It means that $p_i \leq k$ for all $i$. After simple calculation we get the statement of our theorem.

*Remark 2.7.2* Let $A = \{A_{ij}\}$ be the matrix of quadratic form. Let us introduce a formal series

$$\psi_A(q) = \sum_{n_1, \ldots, n_k \geq 0} \frac{q^{\frac{1}{2} \sum_{i,j} A_{ij} n_i n_j}}{(q)_{n_1} \cdots (q)_{n_k}}.$$

The convenient way to rewrite the statement of Theorem 2.7.1 is: $ch\,W(q, 1) = \psi_{B^{-1}}(q)$, where $B^{-1}$ is inverse to the $k \times k$ Cartan matrix

$$B = \begin{pmatrix} 2 & -1 & \cdots & & & 0 \\ -1 & 2 & -1 & & & \\ & -1 & \ddots & & & \\ & & \ddots & & -1 & \\ & & & \ddots & 2 & \\ 0 & & & & -1 & 1 \end{pmatrix}$$

which corresponds to Lie algebra $O_{2k+1}$.

*Remark 2.7.3* Here we present another approach to the left hand side of the Gordon identity. Change the notations a little. The problem is to determine the character of a quotient $C_k = \mathbb{C}[x_1, x_2, \ldots]/(S_m^{(k+1)})$, where $S_m^{(k+1)} = \sum x_{i_1} x_{i_2} \ldots x_{i_{k+1}}$, $i_1 + i_2 + \cdots + i_k = m$. In the algebra $C = \mathbb{C}[x_1, x_2, \ldots]$ let us consider the family of ideals $J^{(k)}$, $k \in \mathbb{Z}$, $k \geq 0$, $J^{(k)}$ is generated by $\{S_m^{(k+1)}\}$. Form the adjoint object: $C/J^{(1)} \oplus J^{(1)}/J^{(2)} \oplus J^{(2)}/J^{(3)} \oplus \cdots = C^{ad}$, $C^{ad}$ is a graded algebra, put $\deg J^{(S)}/J^{(S+1)} = S$. In $C^{ad}$ the images of the elements $S_m^{(k)}$

*(we denote them by the same symbol) form a system of generators. It is possible to prove that they satisfy quadratic relations. Consider two examples:*

(1) $\sum S_m^{(1)} S_n^{(2)} = 0, m+n = a$ in $C^{ad}$. *(This is evident.)*

(2) $\sum_{m+n=a} m S_m^{(1)} S_n^{(2)} = \sum_{m+u+v=a} m x_m x_u x_v = \frac{1}{3} a \sum_{m+u+v=a} x_m x_u x_v = 0 (in\ C^{ad})$.

Similar calculations shows that $\{S_m^{p_1}\}$ and $\{S_n^{p_2}\}$ satisfy $2p_1$ series of relations if $p_1 \leq p_2$. The convenient way to describe $C^{ad}$ again uses the dual coalgebra. Let $V^{(k)}$ be a space with the basis $\{S_k^{(k)}, S_{k+1}^{(k)}, S_{k+2}^{(k)}, \ldots\}$. We identify $(V^{(k)})^* = \Gamma^{(k)}$ with the space of tensors on a line $f(z)(dz)^k$ where $f$ is a polynomial, $\{S_m^{(k)}\}$ and $\{z^m (dz)^k\}$ are the dual bases. Now let us construct a coalgebra $B$. First define $\widetilde{B}$ as:

$$\widetilde{B} = \bigoplus_{i_1, i_2, \ldots, i_r} S^{i_1} \Gamma^{(1)} \otimes S^{i_2} \Gamma^{(2)} \otimes \cdots \otimes S^{i_r} \Gamma^{(r)}, i_\alpha, r \in \mathbb{Z},\ r > 0,\ i_\alpha \geq 0.$$

Elements of $\widetilde{B}$ are expressions:

$$f(z_1(1), z_2(1), \ldots, z_{i_1}(1), z_1(2), \ldots, z_{i_2}(2), \ldots, z_1(r), \ldots, z_{i_r}(r))$$
$$dz_1(1) dz_2(2) \ldots \ldots dz_{i_1}(1) (dz_1(2))^2 (dz_2(2))^2 \ldots (dz_{i_2}(2))^2 \ldots (dz_1(r))^r$$
$$\ldots (dz_{i_r}(r))^r,$$

where $f$ is a polynomial, which is symmetric in the group of variables $z_1(a), z_2(a), \ldots, z_{i_a}(a)$ for each $a$. $\widetilde{B}$ has a natural structure of cocomutative coalgebra. For example, on the component $S^2 \Gamma^{(1)}$ the comultiplication $\widetilde{B} \to \widetilde{B} \otimes \widetilde{B}$ is the natural embedding $S^2 \Gamma^{(1)} \to \Gamma^{(1)} \otimes \Gamma^{(1)}$ plus trivial term $S^2 \Gamma^{(1)} \to S^2 \Gamma^{(1)} \otimes S^0 \Gamma^{(1)} + S^0 \Gamma^{(1)} \otimes S^2 \Gamma^{(1)}$. The coalgebra $B$ is a subcoalgebra in $\widetilde{B}$, which consist of such elements, where the polynomial $f$ has zeros of order $2\min(a,b)$ when $z_i(a) = z_j(b)$. We claim that $B$ is isomorphic to the coalgebra $(C^{ad})^*$. If we work with $C_k$, the description is similar: the coalgebra $(C_k^{ad})^*$ is isomorphic to the subcoalgebra in $B$, which consists of such polynomials $f$, depend only on $z_i(a)$, $a \leq k$. The calculation of the character is the same as above. Let us formulate now the counterpart of the theorem (2.4.1).

**Theorem 2.4.1'** *Irreducible representation of $\widehat{sl}_2$ with highest weight $(0, l, k)$ is realized in the quotient $\widetilde{V}/(S_i^{(k+1)})\widetilde{V}$, where $\widetilde{V}$ is a vector space with the basic of "infinite monomials" $e_{i_1} e_{i_2} \ldots e_{2N}^l e_{2N+1}^{k-l} e_{2N+2}^l e_{2N+3}^{k-l} \ldots v_{-\infty}$ and (1) $e_i$ and $e_j$ can be transposed (they commute) (2) if a monomial m contains fragments $e_i e_{2N}^l e_{2N+1}^{k-l} e_{2N+2}^l \ldots v_{-\infty}$, $i \geq 2N$ or $e_i e_{2N+1}^{k-l} e_{2N+2}^l, \ldots, i \geq 2N$ then $m = 0$. $S_i^{(k)} = \sum e_{\alpha_1} e_{\alpha_2} \ldots e_{\alpha_k}$, $i \in \mathbb{Z}$, $\alpha_1, \ldots, \alpha_k \in \mathbb{Z}$, $\alpha_1 + \cdots + \alpha_k = i$. Elements $e_i \in \widehat{sl}_2$ act by multiplication from the left.*

In the space of representation we can choose the basis from the *basic* monomials.

**Proposition 2.6.1'** *Let us call the monomial $e_{i_1} e_{i_2} \ldots e_{i_m} v$ (or infinite "monomial" $e_{i_1} e_{i_2} e_{i_3} \ldots v_{-\infty}$) basic if  (1) $i_1 \leq i_2 \leq i_3 \leq \cdots$ and in finite case $i_{m-k+l} < -1$ and $i_{j+k} - i_j \geq 2$  (2) if "infinite" case $i_{j+k} - i_j \geq 2$ for all $j$ and the sequence $\{i_\alpha\}$ stabilizes. It means that for some $M$ the structure of sequence is: $i_M = i_{M+1} = \cdots = i_{M+l-1}$, $i_M + 1 = i_{M+l} = i_{M+l+1} = \cdots = i_{M+k}$ and so on. The basic monomials constitute the bases in $W$ and $V$.*

*Remark.* We can certainly formulate the combinatorial version of these statements in terms of statistical sums on a 1-dimensional lattice. For example, the configuration will be a function $f : \mathbb{Z} \to \{0, 1, \ldots, k\}$, such that  (1) $f(x) = 0$ if $x < M$ ($M$ depends on $f$)  (2) $f(x) + f(x+1) \leq k$  (3) if $x > N$ ($N$ again depends on $f$) then $f(x) = l$ if $x$ is even and $f(x) = k - l$ if $x$ is odd.

Then we can define the vacuum configuration $f_{vac}$, $f_{vac}(x) = 0$ if $x < 0$ and $f_{vac}(2y) = l$, $f_{vac}(2y + 1) = k - l$, $y \in \mathbb{Z}$, $y \geq 0$. For each $f$ the difference $S(f) = \sum_{n \in \mathbb{Z}} f(n) - \sum_{n \in \mathbb{Z}} f_{vac}(n)$ if well defined. So, the statistical sum $\sum_f q^{S(f)}$ coincides with the character of the irreducible representation $\widehat{sl}_2$ with highest weight $(0, l, k)$.

Now we can compute the character of $V$. The strategy is the same as for the fundamental representation: the character of the subspace $W$ is given by the left hand side of the Gordon identity and we have to do the "limit" procedure. Put $ch(T^N W) = ch_N$, and then tends $N$ to $-\infty$, $ch(V) = ch_{-\infty}$. We get the formula:

$$(2.6.2') \qquad ch\, V_{(0,l,k)} = \frac{1}{(q)_\infty} \sum_{N_1 \geq N_2 \geq \cdots \geq N_k \in \mathbb{Z}} \frac{q^{N_1^2 + \cdots + N_k^2 + N_{k-l+1} + \cdots + N_k z N_1 + \cdots + N_k + l/2}}{(q)_{N_1 - N_2} (q)_{N_2 - N_3} \cdots (q)_{N_{k-1} - N_k}}$$

*Remark 2.7.4 This formula for the character actually coincides with the "parafermionic" formula due to Lepowsky and Primc [1]. Let us restrict ourselves with the case $l = 0$. Denote by $V(-i, 2j)$ the eigensubspace in $V$, where the energy is equal to $-i$ and the eigenvalue of $h_0$ is equal to $2j$. Suppose first that $j = 0$. We have:*

$$\sum_{i=0}^\infty \dim V(-i, 0) q^i$$
$$= \frac{1}{(q)_\infty} \sum_{\substack{N_1 \geq N_2 \geq \cdots \geq N_k \in \mathbb{Z} \\ N_1 + \cdots + N_k = 0}} \frac{q^{N_1^2 + N_2^2 + \cdots + N_k^2}}{(q)_{N_1 - N_2}(q)_{N_2 - N_3} \cdots (q)_{N_k - N_{k-1}}}$$
$$= \frac{1}{(q)_\infty} \sum_{(n_1, \ldots, n_{k-1})} \frac{q^{\sum B_{ij} n_i n_j}}{(q)_{n_1}(q)_{n_2} \cdots (q)_{n_{k-1}}}, \quad n_i \in \mathbb{Z},\ n_i \geq 0$$

*where $\{B_{ij}\}$ is proportional to the inverse matrix of the Cartan matrix of $A_{k-1}$.*

The general formula is the following: (we use the notations of the remark 2.7.2)

$$\sum_{j=0}^{k-1}\sum_{i=0}^{\infty} \dim V(-i, 2j) q^{i+\frac{j}{k-1}} = \psi_{(A_{k-1}^{-1})}(q) \frac{1}{(q)_\infty}.$$

This is exactly the result of Lepowsky and Primc.

## 3 Flag manifold approach. $sl_2$ case

### 3.1

First fix some notations which concern the flag manifold. Let $G = SL(2, \mathbb{C})$, $\hat{G}$-the central extension of the current group on a circle. Actually $\hat{G}$-one of the possible Kac-Moody groups, but we do not want to be precise at this point. Let $B_+$ be the Borel subgroup in $\hat{G}$ and $P_1$ and $P_2$-two parabolic subgroups of $\hat{G}$. Recall that the Lie algebras of the groups $P_1$ and $P_2$ are the following: $\begin{pmatrix} a & b \\ c & -a \end{pmatrix}$ plus central element, where $a, b, c$ are regular inside the disk and $\begin{pmatrix} a & bz \\ c \cdot z^{-1} & -a \end{pmatrix}$ plus central element where again $a, b, c$ are regular in the disk and $z$ is a local coordinate. The flag manifold is a quotient $F = \hat{G}/B_+$. There are two projections: $\pi_1 : F \to F_1 = \hat{G}/P_1$ and $\pi_2 : F \to F_2 = \hat{G}/P_2$. The fibers of these maps are projective lines. For each point $q \in F$ let us denote by $A_1(q)$ the projective line $\pi_1^{-1}\pi_1(q)$ and $A_2(q)$ is $\pi_2^{-1}\pi_2(q)$. The arrows $q \to A_i(q)$ define two correspondences $A_1, A_2$ on $F$. The Shubert cells (finite dimensional) are the orbits of $B_+$ on $F$. There is a distinguished one-point orbit, we shall denote it by 1. Other orbits correspond to other elements of the affine Weyl group. This Weyl group $W_{\text{aff}}$ is generated by two simple reflections $S_0$ and $S_{-1}$. Each $w \in W_{\text{aff}}$ is a product of them, for example $w = S_0 S_{-1} S_0$, then the closure of the corresponding Shubert cell $\overline{Sh_w} = A_1 A_2 A_1(1)$. It means that each $S_0$ we replace by the correspondence $A_1$ and $S_{-1}$ by $A_2$. Geometrically we start from the point 1 then draw a line in the direction $A_1$, then from each point of this line draw the line in the direction $A_2$, and so on.

Each integrable irreducible representation of $\hat{G}$ is realized in the space of sections of some line bundle $L_\lambda$ on $F$, $H^0(F, L_\lambda) \cong V_\lambda$, $\lambda$ is the highest weight. The important property of Shubert cells is that $H^i(\overline{Sh_w}, L_\lambda) = 0$, $i > 0$, and the natural map $H^0(F, L_\lambda) \to H^0(\overline{Sh_w}, L_\lambda)$ is a surjection.

Now let us introduce the *principal manifold* $M = \overline{\hat{N} \cdot 1} \subset F$-the closure of the orbit of the point 1 with respect to the action of a group of currents with values in the maximal nilpotent subgroup $N$ in $SL(2)$. The Lie algebra of $N$ is $\{e\}$. Manifold $M$ is an inductive limit of finite-dimensional manifolds $M_n$, $M = \varinjlim M_n$ and $M_n = \overline{B_+^n 1}$, $B_n^+ = T^n B_+ T^{-n}$.

It is clear that $M_n$ is isomorphic to one of the Shubert cells (as an orbit of adjoint group). Actually, the following is true. To each $w \in W_{\text{aff}}$ corresponds a point of $F$-the projection of $w$, we denote the image by the same letter. We can construct "Shubert cells" starting from an arbitrary point and acting by correspondence $A_1$ and $A_2$. So $M_n$ is a "Shubert cell", which is constructed by a sequence of correspondences with the starting point $T^n$. As a result we have: $\dim H^i(M, L_\lambda) = 0$ if $i > 0$ and the map $H^0(F, L_\lambda) \to H^0(M, L_\lambda)$ is a surjection. Let us consider the dual map $\varphi : (H^0(M, L_\lambda))^* \to (H^0(F, L_\lambda))^*$. The dual space $(H^0(F, L_\lambda))^*$ is isomorphic to the contragredient representation which is also isomorphic to $L_\lambda$, the map $\varphi$ is an injection and the image is isomorphic to our principal space $W = U(\hat{n})v$. Thus we can use geometric methods to determine the character of $W$. We want to apply the Atiyah-Bott fixed point theorem to the pair $(M, L_\lambda)$, where the maximal abelian subgroup in $\widehat{sl_2}$ acts. On $M$ there is an action of the product of the Cartan subgroup and $\mathbb{C}^*$, this additional $\mathbb{C}^*$ corresponds to the energy. Note that the manifolds $M_n$ are singular, but $M$ is non-singular (see below). The fixed points of the action of our abelian subgroup $\mathfrak{A}$ are $W_{\text{aff}} \cap M$. The fixed point theorem gives us:

$$(3.1.1) \qquad chW^* = \sum_{w \in W_{\text{aff}} \cap M} \frac{e^{w\lambda}}{\sqcap (1 - e^\mu)}$$

where in the product $\sqcap_{\{\mu\}}(1 - e^\mu)$ $\{\mu\}$ is a set of characters of $\mathfrak{A}$ (weights) which appears in the decomposition $T_w^*(M) = \oplus \mu$, $T_w^*(M)$-the tangent space of $M$ in $w$.

## 3.2 The structure of the manifold $M$

**Theorem 3.2.1** *1) $M$ is nonsingular*
*2) $M \cap W_{\text{aff}} = \{T^n : n \geq 0, S_n : n > 0\}$*
*3) The set of all weights which correspond to the action of $\mathfrak{A}$ in the tangent space to $M$ at the point $w \in W_{\text{aff}} \cap M$ is a subset in the root system of $\widehat{sl_2}$. The corresponding root vectors are:*

$$\text{for} \quad w = T^n, n \geq 0 : \{e_{-i}, i \geq 2n+1; f_i, n+1 \leq i \leq 2n; h_{-i}, 1 \leq i \leq n\}$$
$$\text{for} \quad w = S_n, n > 0; \{e_{-i}, i \geq 2n; f_i, n \leq i \leq 2n-1; h_{-i}, 1 \leq i \leq n-1\}$$

*4) The flag manifold $F$ can be decomposed into the union of Shubert cells of finite codimension. They are the orbits of the opposite maximal nilpotent subgroup $N_-$ and are labelled by elements of $W_{\text{aff}} : Sh_w^{cofin} = N_- w, w \in F$. The intersections $Y_w = Sh_w^{cofin} \cap M$ constitute a stratification of $M$ such that*
*(a) $Y_w$ is a contractable complex manifold, $\text{codim} Y_{T_n} = \text{codim} Y_{S_n} = n$*

(b) $Y_w$ are stable with respect to the action of $\hat{N}$, and $Y_{T^n}$ and $Y_{S_n}$ are n-parametric families of the orbits of $\hat{N}$ and the transversal submanifolds are given by formulas:

$$(d_1, \cdots, d_n) \to \begin{pmatrix} (1 + d_1 z^{-1} + \cdots + d_n z^{-n})^{-1} & 0 \\ 0 & 1 + d_1 z^{-1} + \cdots + d_n z^{-n} \end{pmatrix}$$

So, $\{d_1, \cdots, d_n\}$-are the coordinates on the manifold of orbits

(c) The closure of $\bar{Y}_{T^n} = \cup_{m \geq n} Y_{T^m} \cup_{m > n} Y_{S_m}$ and $\bar{Y}_{S_n} = \cup_{m \geq n} Y_{S_m} \cup_{m \geq n} Y'_{T^m}$ where $Y'_{T^m}$ is a subfamily in the set of $\hat{N}$-orbits in $Y_{T^m}$ of codimension one, which consists of the orbits with coordinate $d_m = 0$.

This theorem is proved by a strightford calculation "in coordinates". First let us fix the representation of elements from Weyl group in $\widehat{SL(2)}$, put $T^n = \begin{pmatrix} z^{-n} & 0 \\ 0 & z^n \end{pmatrix}$ and $S_n = \begin{pmatrix} 0 & -z^{-n} \\ z^n & 0 \end{pmatrix}$. The manifold $F$ is covered by the family of coordinate charts $F = \cup_{w \in W_{\text{aff}}} U_w, U_w = wN_- \cdot 1 \cong N_-$. Let us determine, for example, the tangent space $T_w, w = T^1$.

Fix an element $w = T^n = \begin{pmatrix} z^{-n} & 0 \\ 0 & z^n \end{pmatrix}$. Then $w \begin{pmatrix} a & b \\ c & d \end{pmatrix} \in wN_- \cdot 1$, $a = 1 + a_1 z^{-1} + a_2 z^{-2} + \cdots$, $b = b_1 z^{-1} + b_2 z^{-2} + \cdots$, $c = c_0 + c_1 z^{-1} + \cdots$, $d = 1 + d_1 z^{-1} + \cdots$, $ad - bc = 1$ belongs to $\hat{N} \cdot 1$, if and only if for some series $p = p_1 z^{-1} + p_2 z^{-2} + \cdots$ the matrix

$$(3.2.2) \quad \begin{pmatrix} 1 & p \\ 0 & 1 \end{pmatrix} \begin{pmatrix} z^{-n} & 0 \\ 0 & z^n \end{pmatrix} \begin{pmatrix} a & b \\ c & d \end{pmatrix} = \begin{pmatrix} az^{-n} + cz^n p & bz^{-n} + dz^n p \\ cz^n & dz^n \end{pmatrix}$$

is from $B_+$. If $n < 0$, then the series $dt^n = t^n + d_1 t^{n-1} + \cdots$ is not in $\mathbb{C}[z]$, therefore $\hat{N} \cdot 1 \cap U_w = \phi$. Now consider the case $n = 1$. The condition "matrix (3.2.2) belongs to $B_+^{n}$" means in this case that:

$$c = c_0, \quad d = 1 + d_1 z^{-1}$$
$$z^{-1} + a_1 z^{-2} + a_2 z^{-3} + \cdots + c_0(p_1 + p_2 z^{-1} + \cdots) = c_0 p_1,$$
$$b_1 z^{-2} + b_2 z^{-3} + \cdots + (z + d_1)(p_1 z^{-1} + p_2 z^{-2} + \cdots) = p_1.$$

If $c_0 = 0$ then the third equation can not be solved and $p$ does not exist. If $c_0 \neq 0$ then we can find all $p_i, i = 2, 3, \cdots$, for example $p_2 = -c_0^{-1}$. The forth equation gives us $p_1 d_1 + p_2 = 0$ (this is a coefficient before $t^{-1}$ so $p_1 d_1 = c_0^{-1}$ and so it is necessary that $d_1 \neq 0$. Conversely, if $c_0 \neq 0, d_1 \neq 0$, then we can define $p_i$ by the formulas:

$$(3.2.3) \quad p_1 = \frac{1}{c_0 d_1}, \quad p_2 = -\frac{1}{c_0}, \quad p_i = -\frac{a_{i-2}}{c_0} \quad \text{if} \quad i = 3, 4, \cdots$$

The matrix 3.2.2 has the form $\begin{pmatrix} c_0 p_1 & x \\ c_0 z & d_1 + z \end{pmatrix}$ and the determinant is equal to $1, c_0 p_1 (d_1 + z) - x c_0 z = 1$, it means that $x = p_1$ and the matrix is in $B_+$. We obtain the following result:

$$\hat{N} \cdot 1 \cap U_{T^1} = \{T^1 \cdot \begin{pmatrix} 1 + a_1 z^{-1} + \cdots & b_1 z^{-1} + \cdots \\ 0 & 1 + d_1 z^{-1} \end{pmatrix} \cdot 1 : c_0, d_1 \neq 0\}$$

$$M \cap U_{T^1} = \{T^1 \cdot \begin{pmatrix} a & b_1 z^{-1} + \cdots \\ c_0 & 1 + d_1 z^{-1} \end{pmatrix} \cdot 1\} = \{\begin{pmatrix} a & b_1 z^{-3} + b_2 z^{-4} + \cdots \\ c_0 z^2 & 1 + d_1 z^{-1} \end{pmatrix} T^1\}$$

In particular, $M \cap U_{T^1}$ is nonsingular and we see what the structure of tangent space in $T^1$ is. It is clear also why the imaginary roots appear in the tangent space. The hyperplane $\{c_0 = 0\} \subset M \cap U_{T^1}$ is

$$Y_{T^1} = \{\begin{pmatrix} (1 + d_1 z^{-1})^{-1} & b_1 z^{-3} + b_2 z^{-4} + \cdots \\ 0 & 1 + d_1 z \end{pmatrix} T^1\}$$

$$= \cup_{d_1 \in \mathbb{C}} \hat{N} \begin{pmatrix} (1 + d_1 z^{-1})^{-1} & 0 \\ 0 & 1 + d_1 z^{-1} \end{pmatrix} T$$

The hyperplane $\{d_1 = 0\}$ is the intersection $U_{T^1} \cup \bar{Y}_{S_1}$. It can be shown in a similar way.

Let us draw the picture which illustrates the Theorem 3.2.1.

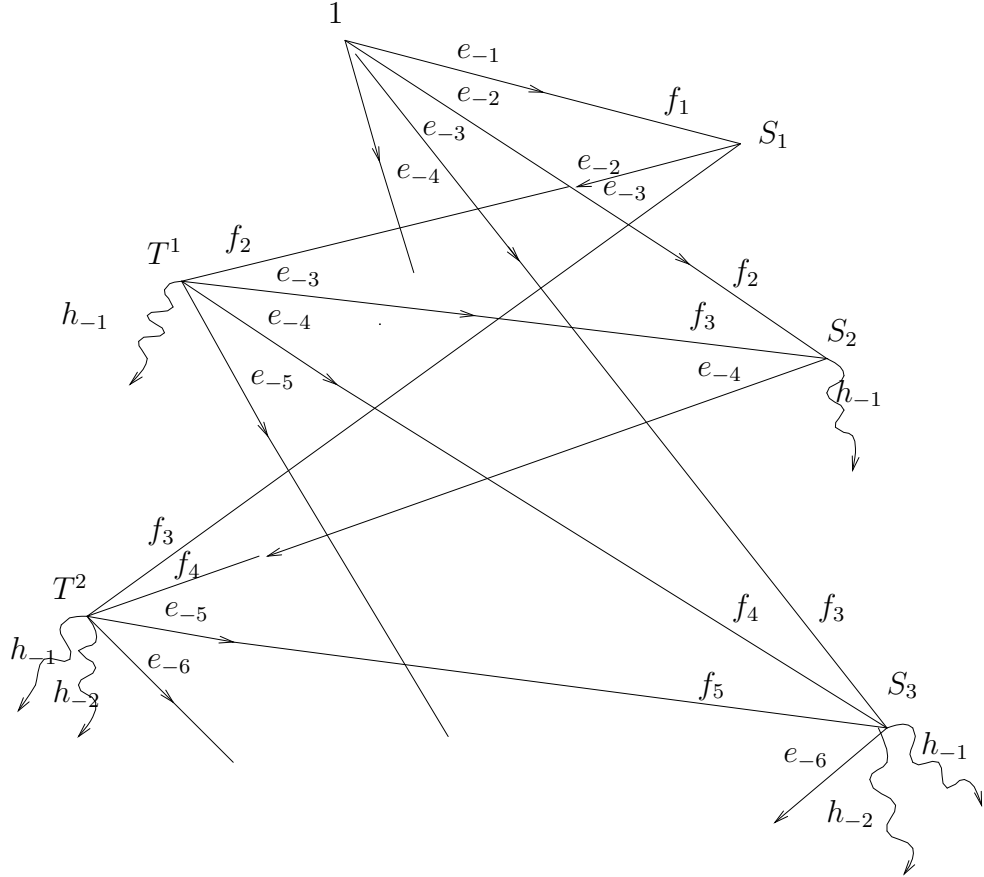

Here the points are elements of the Weyl group, which are identified with the fixed points of $\mathfrak{A}$ on $M$. Consider the Lie subalgebra $sl_2 = \{e_{-2}, f_2, h_0 - 2C\} \subset \widehat{sl}_2$ and let $SL_2 \subset \widehat{SL}_2$ be the corresponding subgroup in $\widehat{SL}_2$. The set $SL_2 \cdot 1 \subset F$ is isomorphic to $\mathbb{CP}^1$-the flag manifold of $SL_2$. The point $S_2$ is in this $\mathbb{CP}^1$. The arrow on the picture, which starts at 1 and ends at $S_2$, is this $\mathbb{CP}^1$. The tangent vectors to $\mathbb{CP}^1$ correspond to $e_{-2}$ in $T_w, w = T^1$ and $f_2$ in $T_w, w = S_2$. The arrows $\rightsquigarrow$ denote the "imaginary" directions.

## 3.3 Combinatorial consequences of Theorem 3.2.1

Let us write down what fixed point formula gives in our case:

(3.3.1)
$$\begin{aligned} ch\, W &= \sum_{n=0}^{\infty} \frac{e^{T^n \lambda}}{(q)_n (1-(q^{n+1}z)^{-1})(1-(q^{n+2}z)^{-1})\ldots(1-(q^{2n}z)^{-1})(q^{2n}z;q)_\infty} + \\ &+ \sum_{n=1}^{\infty} \frac{e^{W_n \lambda}}{(q)_{n-1}(1-(q^n z)^{-1})(1-(q^{n+1}z)^{-1})\ldots(1-(q^{2n-1}z)^{-1})(q^{2n-1}z;q)_\infty} = \\ &= \frac{1}{(z;q)_\infty} \cdot \sum_{n=0}^{\infty} (-1)^n \frac{(z;q)_n}{(q)_n} q^{\frac{3n^2+n}{2}} z^n (e^{T^n\lambda} - e^{S_{n+1}\lambda} z q^{2n+1}) \end{aligned}$$

Here we use the standard notations

$$(a;q)_\infty = \prod_{i=1}^{\infty}(1-aq^i);\ (a;q)_n = \prod_{i=1}^{n}(1-aq^i)$$

Now let us compare different formulas for the character of $W$-(3.3.1) and the Gordon type formula (2.3.3').

**Theorem 3.3.2**

(a) $(1-qz)(1-q^2z)(1-q^3z)\cdots = \sum_{n=0}^{\infty}(-1)^n q^{\frac{3n^2+n}{2}} z^n \frac{(1-qz)\cdots(1-q^n z)}{(1-q)\cdots(1-q^n)}(1-q^{2n+1}z)$

(b) $\sum_{n=0}^{\infty} \frac{q^{n^2} z^n}{(q)_n} = \frac{1}{\prod_{m=1}^{\infty}(1-q^m z)} \sum_{n=0}^{\infty}(-1)^n \frac{(1-qz)\cdots(1-q^n z)}{(1-q)\cdots(1-q^n)}(q^{\frac{5n^2+n}{2}} z^{2n} - q^{\frac{5(n+1)^2-n-1}{2}} z^{2n+1})$

(c) let $k, l \in \mathbb{Z}$, $0 \leq l \leq k$

$$\sum_{N_1 \geq N_2 \geq \cdots \geq N_k \geq 0} \frac{q^{N_1^2 + \cdots + N_k^2 + N_{k-l+1} + \cdots + N_k} z^{N_1 + \cdots + N_k + l/2}}{(q)_{N_1 - N_2}(q)_{N_2 - N_3} \cdots (q)_{N_{k-1} - N_k}(q)_{N_k}} =$$
$$= \frac{1}{\prod_{m=1}^{\infty}(1-q^m z)} \sum_{n=0}^{\infty}(-1)^n \frac{(1-qz)\cdots(1-q^n z)}{(1-q)\cdots(1-q^n)}$$
$$\cdot (q^{\frac{(2k+3)n^2 + (2l+1)n}{2}} z^{(k+1)n + l/2} - q^{\frac{(2k+3)(n+1)^2 - (2l+1)(n+1)}{2}} z^{(k+1)(n+1) - l/2})$$

Point (a) corresponds to the weight $\lambda = (0,0,0)$, (b) to the weight $\lambda = (0,0,1)$ and in (c) we have general case $\lambda = (0,l,k)$.

Specialization $z = 1$ gives us a new proof of the classical partition identities.

**Theorem 3.3.3**

(a) Euler pentagonal theorem: $\prod_{m=1}^{\infty}(1-q^m) = \sum_{n\in\mathbb{Z}}(-1)^n q^{\frac{3n^2+n}{2}}$;

(b) Roger-Ramanujan identities

(I) $\sum_{n=0}^{\infty} \frac{q^{n^2}}{(q)_n} = \prod_{m=1}^{\infty} \frac{1}{(1-q^{5m+1})(1-q^{5m+4})}$

(II) $\sum_{n=0}^{\infty} \frac{q^{n^2+n}}{(q)_n} = \prod_{m=1}^{\infty} \frac{1}{(1-q^{5m+2})(1-q^{5m+3})}$

(c) Gordon identities:

$$\sum_{N_1\geq\cdots\geq N_k\geq 0} \frac{q^{N_1^2+\cdots+N_k^2+N_{k-l+1}+\cdots+N_k}}{(q)_{N_1-N_2}\cdots(q)_{N_{k-1}-N_k}(q)_{N_k}}$$
$$= \prod_{\substack{m\neq 0, \pm(k-l+1)\bmod(2k+3) \\ m>0}} \frac{1}{1-q^m}$$

To get the Gordon identity we have to rewrite the right hand side of (c) from (3.3.2) ($z = 1$). To do it we use Jacobi identity:

$$\sum_{n\in\mathbb{Z}}(-1)^n u^{\frac{n(n+1)}{2}} v^{-n} = (1-v)\prod_{m=1}^{\infty}(1-u^m v^{-1})(1-u^m)(1-u^m v)$$

where $u = q^{2k+3}, v = q^{k-l+1}$

## 3.4

The decomposition $M = UY_w, w \in W_{\text{aff}} \cap M$ can not be used as the Shubert cell decomposition of $F$. The reason is that the closure of $Y_{S_n}$ is not a union of $Y_w$. To improve the situation we have to construct more subtle decomposition. For example, it is necessary to decompose $Y_{T^1} = (Y_{T^1}\backslash Y'_{T^1}) \cup Y'_{T'}$. Suppose we construct the subtle stratification with good properties, then we can use it to get the resolution of $\hat{n}$ module $H^0(M, L_\lambda)$, analogous to the BGG resolution. The terms of this resolution are distributions with support on the cells of decomposition. Character formula shows us that this resolution can not be simple, so the structure of decomposition is complicated. Nevertheless it is possible to find out the initial terms of cell decomposition - cells of codimension zero and one. It gives the two first terms of the resolution of $H^0(M, L_\lambda)$. In particular it gives us a proof of the theorem 2.2.1'.

Let $U = Y_1 = \hat{N}\cdot 1$ be an open dense stratum, $U_1 = U_{S_1}\cap M$, and $U_2 = (U_{T^1}\cap M)\backslash\bar{Y}_{S_1}$ - the open vicinities of strata of codimension 1: $Y_{S_1}$ and $Y_{T^1}\backslash Y'_{T'}$; let $U_1^* = U_1\backslash Y_{S_1}$, $U_2^* = U_2\backslash Y_{T^1}$. From the proof of the theorem 3.2.1 we can deduce that $U_1^* = U_1 \cap U$ and $U_2^* = U_2 \cap U$.

Now we fix the line bundle $L_\lambda$, $\lambda = (0, l, k)$ and we shall write $H^0(Z)$ instead of $H^0(Z, L_\lambda)$, ($Z$-submanifold in $F$). Our geometric data define an exact sequence:

$$(3.4.1) \qquad 0 \to H^0(M) \to H^0(U) \to H^0(U_1^*)/H^0(U_1) \oplus H^0(U_2^*)/H^0(U_2)$$

Here are the usual initial terms of the Grothendieck-Cousin complex, which calculates the cohomology of $M$; the third group of the complex are 1-dimensional local cohomologies with support on the strata of codimension one. It is easy to see that $H^0(U)$ is a cofree $\mathbb{C}[e_{-1}, e_{-2}, \cdots]$-module. So, the dual to 3.4.1 sequence has the form:

$$0 \leftarrow W \xleftarrow{\pi} \mathbb{C}[e_{-1}, e_{-2}, \cdots] \xleftarrow{(\varphi_1, \varphi_2)} M_1 \oplus M_2$$

where $\pi$ is natural projection and $M_i = [H^0(U_i^*)/H^0(U_i)]^*$. Theorem 2.2.1' follows from the lemma.

**Lemma 3.4.2** (1) $M_1$ is a free $\mathbb{C}[e_{-1}, e_{-2}, \cdots]$ module of rank 1 with generator $\tau$, $\varphi_1(\tau) = e_{-1}^{k-l+1}$.

(2) $\mathbb{C}[e_i]$-module $M_2$ is generated by the sequence of generators, $\sigma_i$, $i \in \mathbb{Z}$, $\varphi_2(\sigma_i) = S_i^{(k)}$.

The proof is a straightforward calculation which we will present in an extended version of this paper.

### 3.5

Here we discuss the fixed point formula for singular manifolds and Demazure character formula.

First let us recall what Demazure formula is. Let $\mathfrak{G}$ be a Kac-Moody algebra, $T$-weight lattice, $\mathbb{C}[T]$-the group algebra of $T$ and if $x \in T$ then by $e^x$ we denote this element as a generator of $\mathbb{C}[T]$. The Weyl group $W$ acts on $\mathbb{C}[T]$ by automorphisms; if $\alpha$ is a root, then $S_\alpha$ is the corresponding reflection. The Demazur operator $\sum_\alpha$ acts on $\mathbb{C}[T]$ by the formula

$$\sum_{S_\alpha}(e^x) = \frac{e^x}{1 - e^\alpha} + \frac{e^{S_\alpha x}}{1 - e^{-\alpha}}.$$

Now let $V_\lambda$ be an integrable representation of $\mathfrak{G}$ with highest weight $\lambda$, and $v(w)$-the extremal vector which is corresponded to $w \in W$, $n_-$-the maximal nilpotent subalgebra, $n_- v = 0$. Denote by $V_\lambda(w)$ the subspace $U(n_-)v(w) \subset V_\lambda$. The geometric description of $V_\lambda$ is the following: $V_\lambda(w)$ is dual to the space $H^0(\overline{Sh}_w, L_\lambda)$. Here as usual $L_\lambda$ is a line bundle and $\overline{Sh}_w$ is the closure of Shubert cell. Demazur formula is an expression for the character of the space $V_\lambda(w)$. Fix a reduced decomposition $w = S_{\alpha_1} S_{\alpha_2} \cdots S_{\alpha_l}$ and define $\sum_w$ as $\sum_{\alpha_1} \sum_{\alpha_2} \cdots \sum_{\alpha_l}$. It is well known that $\sum_w$ does not depend on the choice of the reduced decomposition. The character $chV_\lambda(w) = \sum_w e^\lambda$. This formula has another form: $chV_\lambda(w) = \tilde{\sum}_w \cdot e^{w\lambda}$, where $\tilde{\sum}_w = \sum_{\sigma_{\gamma_1}} \sum_{\sigma_{\gamma_2}} \cdots \sum_{\sigma_{\gamma_l}}$ where $\sigma_{\gamma_1}, \sigma_{\gamma_2} \cdots, \sigma_{\gamma_l}$ are reflections corresponding to the set of roots $(\gamma_1, \gamma_2, \cdots, \gamma_l)$ such that each subset $(\gamma_{l-s+1}, \gamma_{l-s+2}, \cdots, \gamma_l)$ coincides with the set of positive roots $\{\gamma \mid w(s)\gamma$ is negative $\}$, $w(s) = S_{\alpha_l} S_{\alpha_{l-1}} \cdots S_{\alpha_{l-1+s}}$. We need this form of the Demazure formula, because $V_\lambda(w) = U(n_-)v(w)$, so we have also the formula for the character of the space $U(wn_-w^{-1})v$.

Now let us apply this to the $\mathfrak{G} = \widehat{sl}_2$-case. If we act on the vacuum vector in $V_\lambda$ by the algebra $\mathbb{C}[e_{-1}]$ we get $\sum_{S_1} e^\lambda$. If we act by $\mathbb{C}[e_{-1}, e_{-2}]$ we get $\sum_{S_2}(\sum_{S_1} e^\lambda)$. The general formula is $ch(\mathbb{C}[e_{-1}, e_{-2}, \cdots, e_{-m}]) = \sum_{S_1} \sum_{S_2} \cdots \sum_{S_m}(e^\lambda)$ (Note that here we mean that $\sum_{S_1}$ acts first, then $\sum_{S_2}, \cdots$). After the direct calculation we get:

$$\begin{aligned}
(3.5.1) \quad chH^0(M_n, L_\lambda)^* &= ch(\mathbb{C}[e_{-1}, e_{-2}, \cdots e_{-2n}]v) \\
&= \sum_{m=0}^{2n-1} \frac{e^{T^m \lambda} \binom{2n-1}{m}_q}{(q^{(-2m-1)}z^{-1}; q)_m (q^{2m}z; q)_{2n-m}} + \\
&+ \sum_{m=1}^{2n} \frac{e^{S_m \lambda} \binom{2n-1}{m-1}_q}{(q^{-2m-1}z^{-1}; q)_m (q^{2m-1}z; q)_{2n-m}}
\end{aligned}$$

Here $\binom{2n-1}{j}_q = \frac{(q)_{2n-1}}{(q)_j (q)_{2n-1-j}}$ is a $q$-binomial coefficient.

If $n \to \infty$, then this character formally tends to the formula (3.3.1). From this point of view it is natural to think that formula (3.3.1) for the character of the basic space coincides with the Demazure formula for an "infinite element" $w_0 = \lim_{n \to \infty} T_n$ in the Weyl group. This $w_0$ can be written as an infinite product in two different ways:

$$w_0 = S_1 S_2 S_3 \cdots = \cdots S_0 S_1 S_0 S_1.$$

Let us make a few comments about the connection of this with the fixed point formula.

Let $X$ be a compact complex algebraic manifold (possibly, with singularities), $L$ is a one-dimensional line bundle on $X$ and $T$ - the torus, which acts algebraically on the pair $(X, L)$, and suppose that $T$ has only finite set of fix points. A natural analog of the fixed point formula is:

$$(3.5.2) \quad \sum (-1)^i ch(T, H^i(X, L)) = \sum_{Tx=x} ch(T, O_x(L))$$

where $O_x(L)$ is a formal completion of the space of sections of $L$ in $x$.

Suppose, for example, that the manifold in the infinitesimal vicinity of the fix point $x$ is a complete intersection. It means that $O_x$ is a quotient $\mathbb{C}[t_1, \cdots, t_M]/J$, where $J$ is generated by the regular sequence $f_1, \cdots, f_N$, $N < M$. Suppose also that the action of $T$ can be lifted on $\mathbb{C}[t_1, \cdots, t_M]$ and $f_i$ are homogeneous elements. In this case the local term from 3.5.2 has the form:

$$(3.5.3) \qquad ch(T, O_x(L)) = e^{\nu} \frac{\prod(1 - e^{\mu_j})}{\prod(1 - e^{\lambda_i})}$$

Here $\nu$ is a weight of the action of $T$ on $O_x(L)/m_x O_x(L)$, $m_x$- maximal ideal, which corresponds to $x$; $\mu_1, \cdots, \mu_N$ are weights of $f_i$ and $\lambda_1, \cdots, \lambda_M$ are weights of $t_i$. Formula (3.5.1) shows that in the fixed points of the torus, $M_n$ should be locally a complete intersection, but we did not verify this fact directly.

## 4 $sl_3$-case

### 4.1

The Lie algebra $sl_3$ has three positive roots $\alpha$, $\beta$, and $\gamma = \alpha + \beta$. The corresponding root vectors are $e^1$, $e^2$ and $e^{12}$, $e^{12} = [e^1, e^2]$. The opposite root vectors are $f^1$, $f^2$ and $f^{12}$. The coroots are $h_\alpha = h^1$, $h_\beta = h^2$ and $h_\gamma = h^{12} = h^1 + h^2$. The basis in $\widehat{sl}_3$ is $e_i^1 = e^1 \otimes z^i$, $e_i^2 = e^2 \otimes z^i, \ldots$ and the central element $C$.

We will try to use notations similar to those in §§2-3. In particular, a weight of $\widehat{sl}_3$ is a triple $\lambda = (m, \nu, k)$, where $k$ is the eigenvalue of central element, $m$ is the energy and $\nu$ is the weight of $sl_3$. Let $V$ (or $V_\lambda$) be an irreducible representation of $\widehat{sl}_3$ with highest weight $\lambda$, $\hat{G}$ —Kac-Moody group, corresponding to Lie algebra $\widehat{sl}_3$, $F = \hat{G}/B_+$ —flag manifold and $L_\lambda$ —homogeneous linear bundle on $F$, which corresponds to $\lambda$. The irreducible representation $V$ is identified with $(H^0(F, L_\lambda))^*$.

Affine Weyl group $W_{\text{aff}} = \check{T} \rtimes S_3$ contains the lattice $\check{T} = \text{Hom}(\mathbb{C}^*, T) \subset \mathfrak{h}$, where $T$ is maximal torus in $SL(3)$, and $\mathfrak{h}$ is the Lie algebra of $T$; $S_3$ is the symmetric group which is the Weyl group $W$ of $SL(3)$.

The reflections, which correspond to the roots $(i, \alpha, 0)$, $(i, \beta, 0)$, $(i, \gamma, 0)$ we denote by $S_\alpha^{-i}$, $S_\beta^{-i}$, $S_\gamma^{-i}$. Simple roots are $\alpha = (0, \alpha, 0)$, $\beta$, and $(1, -\gamma, 0)$ and simple reflections are $S_\alpha$, $S_\beta$, and $S_\gamma^1$.

The action of $\xi \in \check{T}$ and $w \in W$ on the space of weights is given by the formulas:

$$(4.1.1) \qquad \begin{aligned} \xi(m, \lambda, k) &= (m - \lambda(\xi) - \frac{1}{2}k\langle\xi, \xi\rangle, \lambda + k\xi^*, k) \\ w(m, \lambda, k) &= (m, w(\lambda), k) \end{aligned}$$

The canonical scalar product on $\mathfrak{h}$ defines the identification $\mathfrak{h} \to \mathfrak{h}^*$ and $\xi^*$ is the image of $\xi$.

## 4.2

Let $V$ be a fundamental representation of $\widehat{sl}_3$, $V = V_\lambda$, $\lambda = (0,0,1)$. As in §2.1, we define the principal space $W$ as $U(\hat{n})v$, where $v$ is vacuum vector, $n$ is the maximal nilpotent subalgebra in $sl_3$ with basis $e^1$, $e^2$, $e^{12}$, and the basis in $\hat{n}$ is $\{e^1_i, e^2_i, e^{12}_i, i \in \mathbb{Z}\}$, $\hat{n}_-$ is subalgebra in $\hat{n}$ with the same basis, but $i < 0$. We are interested in the left ideal $I$ in $U(\hat{n}_-)$ such that $W = (U(\hat{n}_-)/I)v$. The vectors $f_0^1 v$, $f_0^2 v$, $(e^{12}_{-1})^2 v$ are singular vectors in the Verma representation $M_\lambda$ with highest weight $\lambda = (0,0,1)$. Using this, we deduce that the following vectors belong to $I$ : $(e^{12}_{-1})^2$, $ad\, f_0^1((e^{12}_{-1})^2) = \pm(e^2_{-1}e^{12}_{-1} + e^{12}_{-1}e^2_{-1})$; $ad\, f_0^2((e^{12}_{-1})^2) = \pm(e^1_{-1}e^{12}_{-1} + e^{12}_{-1}e^1_{-1})$, and also $(e^2_{-1})^2$ and $(e^1_{-1})^2$, because, for example, $ad\, f_0^1(e^2_{-1}e^{12}_{-1} + e^{12}_{-1}e^2_{-1}) = \pm 2(e^2_{-1})^2$. As in $\widehat{sl}_2$-case let us introduce the operator $L_{-1}$ from the Virasoro algebra, such that $L_{-1}v = 0$. It gives us five series of elements from $I$. In terms of generating functions these relations are:

$$(4.2.1) \qquad (e^1(z))^2, \quad (e^2(z))^2, \quad e^1(z)e^{12}(z), \quad e^2(z)e^{12}(z)$$

**Theorem 4.2.2** *The left ideal $I$ is generated by the coefficients of the expansions of (4.2.1): $R_k = \sum e_i^1 e_j^1$, $i + j = k$, $i, j \in \mathbb{Z}$, $i, j \leqslant -1$, and so on.*

*Remark* 4.2.3 Actually infinite expressions $\tilde{R}_k = \sum e_i^1 e_j^1$, $i + j = k$, $k \in \mathbb{Z}$, $i + j \in \mathbb{Z}$ and others act by zero on $V$.

Now recall the character formula for $V$ in "bosonic" realization:

$$(4.2.4) \qquad ch\, V = \frac{1}{(q)^2_\infty} \sum_{\xi \in \tilde{T}} e^{\xi \lambda} = \frac{1}{(q)^2_\infty} \sum_{a,b \in \mathbb{Z}} q^{a^2 - ab + b^2} z_1^a z_2^b$$

The notations here are the usual ones: $q$ corresponds to the energy operator and $z_1$ and $z_2$ —to the two simple roots of $sl_3$.

The relations of (4.2.4) and the character of the principal subspace $W$ is the same as in the $\widehat{sl}_2$-case. The character of $W$ is given by the formula:

$$(4.2.5) \qquad ch\, W = \sum_{a,b \geq 0} \frac{q^{a^2 - ab + b^2} z_1^a z_2^b}{(q)_a (q)_b}$$

Note that $ch\, W(q, 1, 1) = \Psi_{A_2}(q)$ (we use the notations of 2.7.2). Note that $I$ is a two-sided ideal in $U(\hat{n}_-)$. We present the description of $U(\hat{n}_-)/I$, which is close to the description of the corresponding quotient in the $sl_2$-case. The first complication is that $U(\hat{n}_-)$ is not a commutative algebra, so first let us do its abelianization.

The algebra $U(\hat{n}_-)$ has a standard filtration such that the adjoint algebra is $S(\hat{n}_-)$ —the symmetric algebra of the space $\hat{n}_-$. The ideal $I$ defines an ideal $I^{ad}$ in $S(\hat{n}_-)$, such that $S(\hat{n}_-)/I^{ad}$ is the adjoint algebra to $U(\hat{n}_-)/I$. In particular, they have the same characters. The commutative algebra $S(\hat{n}_-)$ is generated by the images of the elements $e_i^1$, $e_i^2$, $e_i^{12}$, which we denote by the same symbols.

**Proposition 4.2.6** *The ideal $I^{ad}$ is generated by relations (4.2.1).*

This proposition says that we can work with a commutative algebra with quadratic relations. Now following the line of 2.3, we can describe the dual space to the $W^{ad} = S(\hat{n}_-)/I^{ad}$. The space $W^{ad}$ has a $\mathbb{Z}^3$-grading $(n_1, n_2, n_3)$ where $n_1$ is degree with respect to the group of variables $\{e_i^1\}$, $n_2$ —the degree with respect to $\{e_i^2\}$ and $n_3$ corresponds to $\{e_j^{12}\}$. The $(r,s,t)$ component of a $(W^{ad})^*$ is isomorphic (see 2.3.2) to some space of polynomials:

$$(W^{ad})^*_{r,s,t} \cong \{g(x_1,\ldots,x_r; y_1,\ldots,y_s; u_1,\ldots,u_t) \cdot \prod_{i<j}(x_i-x_j)^2 \times$$
$$\times \prod_{i<j}(y_i-y_j)^2 \prod_{i<j}(u_i-u_j)^2 \prod_{i,j}(x_i-y_j)\prod_{i,j}(x_i-u_j)\prod_{i,j}(y_i-u_j)\prod_i dx_i \prod_j dy_j \prod_k du_k\}$$

where $g$ is a polynomial which is symmetric with respect to $\{x_i\}$, $\{y_i\}$, $\{u_i\}$.

Therefore:

$$ch\, W = \sum_{r,s,t} ch(W^{ad})^*_{r,s,t} = \sum_{r,s,t} \frac{q^{r^2+rs+s^2+st+t^2} z_1^{r+s} \cdot z_2^{s+t}}{(q)_r (q)_s (q)_t}$$

Using the simple formulas with $q$-binomial coefficients, it is possible to show that this sum is equal to (4.2.5).

## 4.3 Principal manifold $M$ and fixed point formula

In this section we restrict ourselves only with the "vacuum" representation with highest weight $\lambda = (0,0,k)$, $k \geq 0$, $k \in \mathbb{Z}$. In this case the irreducible representation $V$ is realized in the space of sections of a bundle $L_\lambda$ on a parabolic flag manifold $P = \hat{G}/\hat{G}^{in}$, where the subgroup $\hat{G}^{in}$ corresponds to the subalgebra $\hat{\mathfrak{g}}^{in} = \mathbb{C} \cdot C + \mathfrak{g} + \mathfrak{g}z + \mathfrak{g}z^2 + \cdots$. In other words, $\hat{\mathfrak{g}}^{in}$ consists of currents on the circle which can be extended analytically inside the disc on the complex plane. So we shall work with $P$ and the principal manifold is the closure of the orbit $\hat{N}_- \cdot 1$, where 1 is the zero-dimensional Shubert cell on $P$.

Our problem now is to write down the fixed point formula like (3.3.1) for the pair $(M, L_\lambda)$ and 3-dimensional torus which acts on $(M, L_\lambda)$. The fixed points of the torus in $P$ are labeled by the elements of the lattice $\check{T}$

**Theorem 4.3.1** *1)* $\check{T} \cap M = \{T_{m\alpha+n\beta} : m, n \geqslant 0\}$ *(see picture 4)*

*2) $M$ is nonsingular at the points $T_{n\alpha}$, $T_{n\beta}$ and singular in all others fixed point.*

*3) Local term in the fixed point formula corresponding to $T_{n\alpha}$ is equal to*

$$\Delta_{n\alpha} = \frac{e^{T_{n\alpha}\lambda}}{\sqcap(1-e^\delta)}$$

*where $\delta$ is a root of $\widehat{sl}_3$ such that, $S_\delta(n\alpha) \neq n\alpha$; $T_{n\alpha}(\delta) < 0$, $S_\delta(n\alpha) = k\alpha + l\beta$, $k, l \geq 0$, $S_\delta$ —reflection with respect to $\delta$. In the variables $q, z_1, z_2, \Delta_{n\alpha}$ is given by*

$$\Delta_{n\alpha} = \frac{e^{T_{n\alpha}\lambda}}{(q^{-2n}z_1^{-1};q)_n(q^{2n}z_1;q)_\infty(z_2,q^{-1})_{n-1}(1-z_2)(z_2;q)_\infty(q^n z_1 z_2;q)_\infty}$$

*The local term $\Delta_{n\beta}$ corresponding to $T_{n\beta}$ can be obtained from $\Delta_{n\alpha}$, if we change $\alpha \leftrightarrow \beta$, $z_1 \leftrightarrow z_2$.*

*4) At the points $T_{n\gamma} = T_{n\alpha+n\beta}$ $M$ is locally a complete intersection and*

$$\Delta_{n\gamma} = \frac{e^{T_{n\gamma}\lambda}}{\sqcap(1-e^\delta)} \cdot \frac{(1-(z_1 z_2)^{-1})(1-(qz_1 z_2)^{-1}) \ldots (1-(q^{n-1}z_1 z_2)^{-1})}{(1-q)(1-q^2)\ldots(1-q^n)}$$

*where $S_\delta(-n\gamma) \neq -n\gamma$, $S_\delta(-n\gamma) = k\alpha + l\beta$, $k, l \geq 0$, $T_{n\gamma}(\delta) < 0$.*

$$\sqcap(1-e^\delta) = (1-z_1^{-1})(1-(qz_1)^{-1})\ldots(1-(q^{n-1}z_1)^{-1})(1-q^{n+1}z_1)(1-q^{n+2}z_1)\ldots \cdot$$
$$\cdot (1-z_2^{-1})(1-(qz_2)^{-1})\ldots(1-(q^{n-1}z_2)^{-1})(1-q^{n+1}z_2)(1-q^{n+2}z_2)\ldots \cdot$$
$$\cdot (1-(q^n z_1 z_2)^{-1})(1-(q^{n+1}z_1 z_2)^{-1})\ldots(1-(q^{2n-1}z_1 z_2)^{-1})(1-q^{2n+1}z_1 z_2)(1-q^{2n+2}z_1 z_2)\ldots$$

We could not determine the local terms $\Delta_{n\alpha+m\beta}$, if $n \neq, m \neq 0$ and $n \neq m$. The singularities of $M$ at such points are complicated and we think that $M$ is not a complete intersection in the neighbourhood of such points. Note that Demazure's formula can be used in this situation, but we did not carry out the calculation.

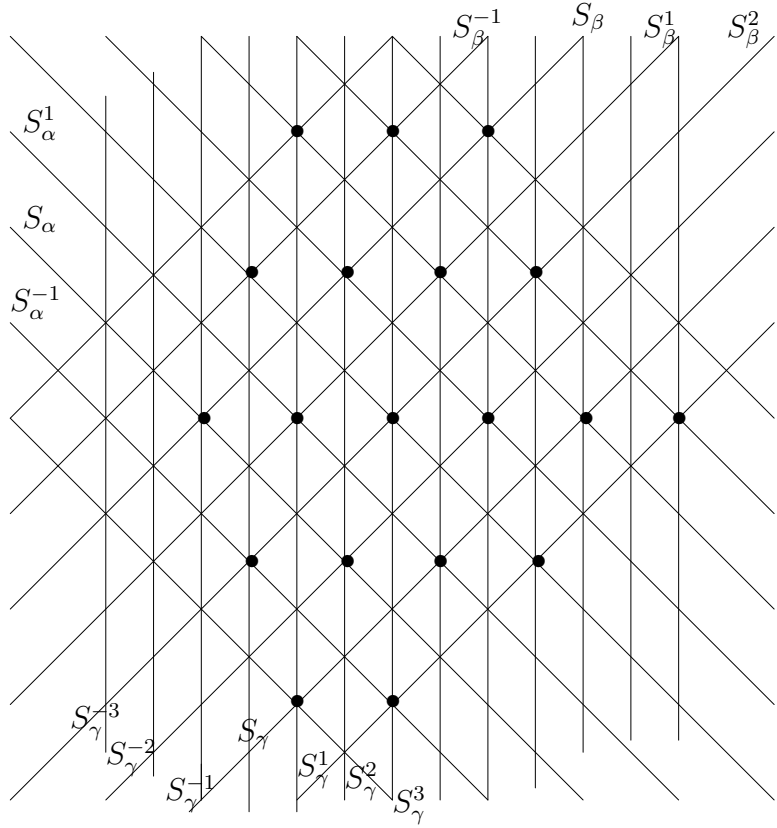

Picture 4

The straight lines on the picture are the mirrors, corresponding to reflections with respect to the affine roots. Distinguished points belong to $\check{T} \cap M$.

**Conjecture 4.3.2** *The local terms $\Delta_{m\alpha+n\beta}$, $m \neq 0$, $n \neq 0$, $m \neq n$ are zero if $z_1 = z_2$. We suppose that such $\Delta_{m\alpha+n\beta}$ can be represented as a product of $(1-z_1)$ or $(1-z_2)$ or $(1-z_1 z_2)$ and something else.*

Let us calculate $\Delta_{n\alpha}$, $\Delta_{n\beta}$, and $\Delta_{n\gamma}$ when $z_1 = z_2 = 1$. Certainly, it can not be done, because each term contains $1-z_1$ or $1-z_2$ or $1-z_1 z_2$ in the denominator. Nevertheless these infinites cancel each other in the sum $\Delta_{n\alpha} + \Delta_{n\beta} + \Delta_{n\gamma}$. The calculation gives the following result.

**Theorem 4.3.3** $(\Delta_{n\alpha} + \Delta_{n\beta} + \Delta_{n\gamma})|_{z_1=z_2=1}$ *is equal*

(a) $\dfrac{(6n+1)q^{3n^2+n} - (6n-1)q^{3n^2-n}}{(q)_\infty^3}$ *if $V$ is fundamental representation ($k=1$).*

(b) $\dfrac{((2k+4)n+1)q^{(k+2)n^2+n} - ((2k+4)n-1)q^{(k+2)n^2-n}}{(q)_\infty^3}$ if $V$ is representation with highest weight $(0,0,k)$.

**Theorem 4.3.4** *(Modulo conjecture 4.3.2)*
(a) *Gauss formula:* $(q)_\infty^3 = 1 - 3q + 5q^3 - 7q^6 \ldots + (-1)^n(2n+1)q^{\frac{n(n+1)}{2}} + \ldots$
(b) *Analog of Roger-Ramanujan:*

$$\sum_{a,b\geq 0} \frac{q^{a^2-ab+b^2}}{(q)_a(q)_b} = \frac{\sum_{n\in\mathbb{Z}}(6n+1)q^{3n^2+n}}{(q)_\infty^3}$$

(c) $$\Psi_{A_2\otimes B_k^{-1}}(q) = \frac{\sum_{n\in\mathbb{Z}}((2k+4)n+1)q^{(k+2)n^2+n}}{(q_\infty)^3}$$

Here (a) corresponds to $k=0$, (b) to $k=1$ and (c) is to arbitrary $k$. The left hand side of the formula (c) is written in the notation of 2.7.2, quadratic form is a product of $B_k^{-1}$ (the same as in Gordon formula) and Cartan matrix for $SL(3)$. We shall discuss this expression in §4.5.

## 4.4

Note that the right hand side of (b) from 4.3.4 coincides with the Kac formula for the character of the fundamental representation of $\widehat{sl}_2$ with highest weight (0,1). And the same is true for (c): the character of $\widehat{sl}_2$ representation with highest weight $(0,k)$ is equal to the right hand side of (c). It is very easy to see for (b). Recall the simple combinatorial fact (it concerns Dufrey square):

$$\sum_{\substack{a,b\geq 0 \\ a-b=m=\text{const}}} \frac{q^{ab}}{(q)_a(q)_b} = \frac{1}{(q)_\infty}$$

Then

$$\sum_{a,b\geq 0} \frac{q^{a^2-ab+b^2}}{(q)_a(q)_b} = \sum_{\substack{m\in\mathbb{Z} \\ a,b, a-b=m}} q^{m^2} \cdot \sum \frac{q^{ab}}{(q)_a(q)_b} = \sum_m \frac{q^{m^2}}{(q)_\infty}$$

And we get the character of fundamental representation of $\widehat{sl}_2$ in "bosonic" realization. We do not know the direct proof of (c).

Now we give some explanation of this fact. Orthogonal subalgebra in $sl_3$ is isomorphic to $sl_2$, it gives us a map $\widehat{sl}_2 \to \widehat{sl}_3$. Restriction of the representation of $\widehat{sl}_3$ of level $k$ on $\widehat{sl}_2$ is a representation of level $k$. So, we have in $\widehat{sl}_3$ two subalgebras $\widehat{sl}_2$ and $\hat{n}$. They have some similarity. Namely, in $sl_3$ there are two 3-dimensional subalgebras: orthogonal and maximal nilpotent and maximal nilpotent is a deformation of orthogonal. $\widehat{sl}_2$ and $\hat{n}$ are currents with value in $sl_2$ and $n$ (and central element add in the first case). In the space of vacuum representation $V_\lambda$, $\lambda = (0,0,k)$ of $\widehat{sl}_3$ there are two subspaces: $U(\hat{n})v = W$ and $U(\widehat{sl}_2)v = W_2$. Energy operator $L_0$ is acting in $W_1$ and $W_2$ and the fact about characters means that $tr\, q^{L_0}|W_1 = tr\, q^{L_0}|W_2$. Now for simplicity restrict ourselves only by the case $k=1$. Space $W_2$ is fundamental representation of $\widehat{sl}_2$ and $W_2 \cong U(\widehat{sl}_2^{out})/I$, where $\widehat{sl}_2^{out}$ is the Lie algebra with the basis $\{e_i, h_i, f_i, i < 0\}$ and $I$ is a left ideal. Five elements $e_{-1}^2$, $[f_0, e_{-1}^2] = -(h_{-1}e_{-1} + e_{-1}h_{-1})$, $[f_0[f_0\, e_{-1}^2]] = -2(f_{-1}e_{-1} + e_{-1}f_{-1}) + h_{-1}^2$, $h_{-1}f_{-1} + f_{-1}h_{-1}$, $f_{-1}^2$ belong to $I$. Operator $L_{-1}$ from Virasoro gives five serieses of generators of $I$:

$$(4.4.1) \quad \begin{array}{l} (1) \sum\limits_{i+j=-n} e_i e_j \quad (2) \sum\limits_{i+j=-n} h_i e_j + e_j h_i \quad (3) \sum\limits_{i+j=-n} f_i e_j + e_i f_j - h_i h_j \\ (4) \sum\limits_{i+j=-n} h_i f_j + f_j h_i \quad (5) \sum\limits_{i+j=-n} f_i f_j \end{array}$$

**Proposition 4.4.2** (a) Five serieses of relations (4.4.1) generate the ideal $I$ (b) the same is true for the adjoint graded space. It means that we replace $U(\widehat{sl}_2^{out})$ by adjoint space of polynomials on $(\widehat{sl}_2^{out})^*$ and $I$ defines the adjoint ideal $I^{ad}$ in $S^*(\widehat{sl}_2^{out})$. Now in $S^*(\widehat{sl}_2^{out})$ the variables $e_i, h_i, f_i$ commute with each other and $I^{ad}$ is generated by the same expression (4.4.1).

Now let us compare 4.4.2 (b) and 4.2.6. We see that the quotient spaces $S(\hat{n}_-)/I^{ad}$ and $S(\widehat{sl}_2^{out})/I^{ad}$ are almost identical: the difference is only in the third series of relations, there is the additional term $\sum h_i h_j$.

These considerations can be generalized to arbitrary $k$.

## 4.5

In this section we give another description of $W = U(\hat{n}_-)/I$ following Schechtman-Varchenko. To do it, first let us describe the universal enveloping algebra $U(\hat{n}_-)$ in terms of the space of symmetric polynomials. Actually, we shall give a new construction of dual coalgebra.

Let $\overline{W} = \oplus \overline{W}_{m,n}$, $m, n \in \mathbb{Z}$, $m, n \geq 0$ be the space of functions:

$$\overline{W}_{m,n} = \{f(x_1, \ldots, x_n; y_1, \ldots y_m\}$$

$f$ is polynomial in $\{x_i\}$, $\{y_j\}$ which is symmetric with respect to the group of variables $\{x_i\}$ and $\{y_j\}$.

We shall call $\{x_i\}$ the coordinates of the particles of type $\alpha$ and $\{y_j\}$ —the coordinates of the particles of type $\beta$. (Recall that $\alpha, \beta$ are positive simple roots of $sl_3$). $\overline{W}$ has a natural structure of commutative and cocommutative Hopf algebra. Multiplication $\overline{W}_{m',n'} \otimes \overline{W}_{m'',n''} \to \overline{W}_{m,n}$, $m = m' + m''$, $n = n' + n''$ is given by the formula:

$$f \cdot g(x_1, \ldots, x_m; y_1, \ldots, y_n) = \text{Symm}\, f(x_1, \ldots, x_{m'}; y_1, \ldots y_{n'}) g(x_{m'+1}, \ldots, x_m; y_{n'+1}, \ldots, y_n),$$

where Symm is the symmetrization with respect to the groups of variables $\{x_i\}$ and $\{y_j\}$. The component of comultiplication $\Delta : \overline{W}_{m,n} \to \overline{W}_{m',n'} \otimes \overline{W}_{m'',n''}$ is nontrivial only if $m = m' + m''$, $n = n' + n''$ and in this case $\Delta$ is a natural embedding. The space $\overline{W}_{m',n'} \otimes \overline{W}_{m'',n''}$ may be identified with some space of polynomials in $\{x_i\}$, $1 \leq i \leq m$, $\{y_j\}$, $1 \leq j \leq n$ with evident symmetry properties and $\Delta(\overline{W}_{m,n}) \subset \overline{W}_{m',n'} \otimes \overline{W}_{m'',n''}$ consists of polynomials with extra symmetry conditions. We now want to extend $\overline{W}$ to $\widetilde{W}$; elements of $\widetilde{W}$ are rational function in $\{x_i\}$, $\{y_j\}$, which can have a pole of order one when $x_i = y_j$. Multiplication makes sense for such functions but comultiplication acts from $\widetilde{W}_{m,n}$ into some modified tensor product $\widetilde{W}_{m',n'} \widehat{\otimes} \widetilde{W}_{m'',n''}$. Symbol $\widehat{\otimes}$ means that functions from the space $\widetilde{W}_{m',n'} \widehat{\otimes} \widetilde{W}_{m'',n''}$ can have poles of order one in the hyperplanes $x_i = y_j$.

Let us define a subspace $A_{m,n}$ in the space of functionals on $\widetilde{W}_{m,n}$. Elements from $\widetilde{W}_{m,n}$ can be represented as products $p = f(x_1, \ldots, x_m, y_1, \ldots y_n) \sqcap (x_i - y_j)^{-1}$, $1 \leq i \leq m$, $1 \leq j \leq n$. For each pair $(i, j)$ fix one of two possible decompositions of $(x_i - y_j)^{-1}$: $\tau_+(x_i, y_j) = x_i(1 + y_j x_i^{-1} + (y_j x_i^{-1})^2 + \ldots)$, $\tau_-(x_i, y_j) = -y_j(1 + (x_i y_j^{-1}) + (x_i y_j^{-1})^2 + \ldots)$. After it we can form the Laurent expansion of $p$. The space $A_{m,n}$ is generated by functionals $a$, such that the pairing $\langle a, p \rangle$ is some coefficient of this Laurent expansion. Then $a$ depends on the choice of the coefficient and the choice of the expansions $\tau_+$ or $\tau_-$ of each term $(x_i - y_j)^{-1}$. $A = \underset{m,n}{\oplus} A_{m,n}$ is a cocommutative Hopf algebra. The comultiplication is evident and multiplication is calculated by the formula:

$$a_1 \in A_{m',n'}, \quad a_2 \in A_{m'',n''}, \quad p \in \widetilde{W}_{m,n}, \quad m = m' + m'', \ n = n' + n'',$$

$$\Delta p \in \widetilde{W}_{m',n'} \widehat{\otimes} \widetilde{W}_{m'',n''}, \quad \Delta p = f(x_1, \ldots, x_m; y_1, \ldots, y_n) \sqcap (x_i - y_j)^{-1},$$

$$\Delta p_+ = f \cdot \underset{\substack{1 \leq i \leq m' \\ 1 \leq j \leq n'}}{\sqcap} (x_i - y_j)^{-1} \cdot \underset{\substack{m'+1 \leq i \leq m \\ n'+1 \leq i \leq n}}{\sqcap} (x_i - y_j)^{-1} \cdot \underset{\substack{1 \leq i \leq m' \\ n'+1 \leq i \leq n}}{\sqcap} \tau_+(x_i, y_j) \cdot \underset{\substack{m'+1 \leq i \leq m \\ 1 \leq j \leq n'}}{\sqcap} \tau_+(x_i, y_j)$$

$$\langle a_1 \cdot a_2, p \rangle = \langle a_1 \otimes a_2, \Delta p_+ \rangle$$

So, to determine the pairing of a product $a_1 \cdot a_2$ on $p$ we have to find $\Delta p$, choose the expansions of all $(x_i - y_j)^{-1}$ and then the value of the pairing is some Laurent coefficients of the resulting series $\Delta p_+$.

Now let us introduce the Serre relations. Define a subcoalgebra $\widetilde{U}$ in $\widetilde{W}$, which consists of expressions $f(x_1, \ldots, x_m; y_1, \ldots, y_n) \sqcap (x_i - y_j)^{-1}$ where the polynomial $f$ is zero, if $x_1 = x_2 = y_1$ or $x_1 = y_1 = y_2$. It is clear that $\widetilde{U}$ is stable with respect to comultiplication, it means that $U$ is a Hopf algebra "in extended sense" (it means that the comultiplication acts to the extended tensor product). It is more convenient for our goals to introduce the subcoalgebra $U \subset \tilde{U}$, $U = \{f(x_1,\ldots,x_m; y_1\ldots,y_n)x_1\ldots x_m\, y_1\ldots y_n \sqcap (x_i - y_j)^{-1}\}$. The dual space $U^*$ (in the same sense as above) is a quotient of $A$. It can be shown that $U^*$ is isomorphic to $U(\hat{n}_-) = U(nz^{-1} + nz^{-2} + \ldots)$, $n$ —the maximal nilpotent subalgebra in $sl_3$. So, we have a nice "functional" model of $(U(\hat{n}_-))^*$. Let us return to the principal space $W_k = U(\hat{n}_-)/I_k$ ($k$ means that we deal with the vacuum representation of level $k$); $W^*$ is subspace in $(U(\hat{n}_-))^*$.

**Theorem 4.5.1** *The dual space to the principal subspace $W_k$ is isomorphic to the space $M = \oplus M_{m,n}$, $m,n \in \mathbb{Z}$, $m,n \geq 0$, where*

$$M_{m,n} = \{f(x_1,\ldots,x_m,y_1,\ldots,y_n)x_1\ldots x_m\, y_1\ldots y_n \sqcap (x_i - y_j)^{-1}\},$$

*$f$ is polynomial and symmetric with respect to $\{x_i\}$, $\{y_j\}$ and $f$ is zero if $x_1 = x_2 = y_1$, $y_1 = y_2 = x_1$, $x_1 = x_2 = \cdots = x_{k+1}$ or $y_1 = y_2 = \cdots = y_{k+1}$. For $k=1$ a simple calculation gives us the formula (4.2.5). For general $k$ we need arguments with filtration like in the $\widehat{sl}_2$ case, where we obtained the left hand side of the Gordon identity.*

$$ch\, W_k(q,z_1,z_2) = \sum_{\substack{0 \leq N_1 \leq N_2 \leq \cdots \leq N_k \\ 0 \leq M_1 \leq M_2 \leq \cdots \leq M_k}} \frac{q^{N_1^2+N_2^2+\cdots+N_k^2+M_1^2+\cdots+M_k^2 - \sum_{i<j} M_i N_j}\, z_1^{\sum N_i}\, z_2^{\sum M_j}}{(q)_{N_1}(q)_{N_2-N_1}\cdots(q)_{N_k-N_{k-1}}(q)_{M_1}(q)_{M_2-M_1}\cdots(q)_{M_k-M_{k-1}}}$$

If $z_1 = z_2 = 1$ this is exactly $\Psi_{A_2 \otimes B_k^{-1}}$ from 4.3.4 (c).

Now we can use the formula for the character of $W_k(q,z_1,z_2)$ to obtain the following formula for the full character of the irreducible representation $\widehat{sl}_3$ with highest weight $(0,0,k)$

$$ch\, V_{(0,0,k)} = \frac{1}{(q)_\infty^2} \cdot \sum_{\substack{N_1 \leq N_2 \leq \cdots \leq N_k \\ M_1 \leq M_2 \leq \cdots \leq M_k}} \frac{q^{N_1^2+N_2^2+\cdots+N_k^2+M_1^2+\cdots+M_k^2 - \sum_{i<j} M_i N_j}\, z_1^{\sum N_i}\, z_2^{\sum M_j}}{(q)_{N_2-N_1}\cdots(q)_{N_k-N_{k-1}}(q)_{M_2-M_1}\cdots(q)_{M_k-M_{k-1}}}$$

$N_i, M_i \in \mathbb{Z}$

## Literature

1. J. Lepowsky and M. Primc, Structure of the standard modules for the affine Lie algebra $A_1^{(1)}$. Contemporary Mathematics V.**46**, AMS 1985.

2. M. Kashiwara, The flag manifold of Kac-Moody Lie algebra. In: Algebraic analysis, geometry, and number theory, edited by Jun-Ichi Igusa.